\documentclass[onefignum,onetabnum]{siamart190516}

\usepackage{lipsum}
\usepackage{amsfonts}
\usepackage{graphicx}
\usepackage{epstopdf}
\usepackage{algorithm}
\usepackage{algpseudocode}
\usepackage{eurosym}
\usepackage{graphicx,amsmath}
\usepackage{setspace}
\usepackage{amssymb}
\usepackage{dsfont}
\usepackage{amsmath}
\usepackage{amsfonts}
\usepackage{ifthen}
\usepackage{mathtools}
\usepackage{enumerate}
\usepackage{enumitem}
\usepackage{color,tikz}
\usepackage{framed}
\usepackage{caption}
\usepackage{subcaption}
\usepackage{cancel,array}
\ifpdf
  \DeclareGraphicsExtensions{.eps,.pdf,.png,.jpg}
\else
  \DeclareGraphicsExtensions{.eps}
\fi

\newsiamremark{remark}{Remark}
\newsiamthm{assumption}{Assumption}
\crefname{assumption}{Assumption}{Assumptions}
\newsiamthm{claim}{Claim}

\headers{Fictitious Play in Zero-sum Stochastic Games}{M. O. Sayin, F. Parise, and A. Ozdaglar}

\title{Fictitious Play in Zero-sum Stochastic Games\thanks{This is the extended arXiv version of \cite{ref:Sayin20} and contains some additional technical details deferred in \cite{ref:Sayin20}. 
\funding{This research was supported by the U.S. Army Research Office (ARO) grant W911NF-18-1-0407.}}}

\author{Muhammed O. Sayin\thanks{Department of Electrical and Electronics Engineering, Bilkent University, Ankara, Turkey, 06800. (\email{sayin@ee.bilkent.edu.tr})}
\and Francesca Parise\thanks{Department of Electrical and Computer Engineering, Cornell University, Ithaca, NY, 14853.
  (\email{fp264@cornell.edu}).}
\and Asuman Ozdaglar\thanks{Department of Electrical Engineering and Computer Science, Massachusetts Institute of Technology, Cambridge, MA, 02139. 
  (\email{asuman@mit.edu}).}}

\usepackage{amsopn}

\ifpdf
\hypersetup{
  pdftitle={Fictitious Play in Zero-sum Stochastic Games},
  pdfauthor={M. O. Sayin, F. Parise, and Asuman Ozdaglar}
}
\fi

\DeclareRobustCommand\dotted{\tikz[baseline=-0.6ex]\draw[ultra  thick,dotted] (0,0)--(0.58,0);}

\DeclareMathOperator*{\argmax}{argmax}
\newcommand{\player}[1]{player $#1$}
\newcommand{\Player}[1]{Player $#1$}
\newcommand{\E}[1]{\mathbb{E}\left\{#1\right\}}
\newcommand{\prob}[1]{\mathbb{P}\left\{#1\right\}}
\newcommand{\indicator}[1]{\mathbb{I}_{\{#1\}}}
\newcommand{\val}{\mathrm{val}}

\newcommand{\as}{\mathrm{a}}
\newcommand{\x}{\mathrm{x}}

\newcommand{\mypi}{\boldsymbol{\pi}}
\newcommand{\Q}{\mathrm{Q}}
\newcommand{\h}{\mathrm{h}}
\newcommand{\oepsilon}{\overline{\epsilon}}
\newcommand{\uepsilon}{\underline{\epsilon}}

\newcommand{\rhoa}{\rho_{\alpha}}
\newcommand{\rhob}{\rho_{\beta}}

\def\mycmd{1}
\def\mymain{0}
\newcommand{\arXiv}[2]{\ifx\mycmd\mymain
#1
\else
#2
\fi}

\begin{document}

\maketitle

\begin{abstract}
We present a novel variant of fictitious play dynamics combining classical fictitious play with $Q$-learning for stochastic games and analyze its convergence properties in two-player zero-sum stochastic games. Our dynamics involves players forming beliefs on the opponent strategy and their own continuation payoff ($Q$-function), and playing a  greedy best response by using the estimated continuation payoffs. Players update their beliefs from observations of opponent actions. A key property of the learning dynamics is that update of the beliefs on $Q$-functions occurs at a slower timescale than update of the beliefs on strategies. We show both in the model-based and model-free cases (without knowledge of player payoff functions and state transition probabilities), the beliefs on strategies converge to a stationary mixed Nash equilibrium of the zero-sum stochastic game.
\end{abstract}

\begin{keywords}
stochastic games, fictitious play, $Q$-learning, two-timescale learning.
\end{keywords}

\begin{AMS}
  91A15, 91A26, 68T05
\end{AMS}

\section{Introduction}
A common justification for Nash equilibrium is that it arises from the learning dynamics of myopic players taking greedy best response actions. This perspective has been investigated for various classes of strategic-form games (also referred to as one-shot or normal-form games) mostly focusing on best-response type dynamics (including fictitious play) \cite{ref:Leslie05,ref:Marden09,ref:Swenson17,ref:Swenson18}. Nevertheless study of learning dynamics in the context of stochastic games has been limited.

Stochastic games were introduced by \cite{ref:Shapley53} to model interactions among multiple players in a multi-state dynamic environment. Players' actions determine not only the payoffs at the current state, {\it stage payoffs}, but also the transition probability to the next state, and hence the continuation payoffs. The decision problem of a player thus involves trading off current  stage payoff for estimated continuation payoffs while forming predictions on the opponent's strategy. This dynamic trade-off makes the analysis of learning in stochastic games potentially challenging.

\subsection{Contributions}

In this paper, we present a novel variant of fictitious play combining classical fictitious play with $Q$-learning \cite{ref:Watkins92} for stochastic games and analyze its convergence properties in two-player zero-sum stochastic games. Each player forms a belief on the opponent's (stationary) mixed strategy and her own {\em $Q$-function}, which corresponds to her  continuation payoff given the opponent's strategy. Players play a greedy best response strategy in an {\em auxiliary game} with player payoffs given by the sum of the stage payoffs and estimated continuation payoffs. At each stage of the game over the infinite horizon, the players update their beliefs on the opponent strategy from the observation of the other player's action. The update of the $Q$-function is then constructed as the maximum payoff that can be attained in the auxiliary game over all possible actions (given beliefs on opponent play and $Q$-function). 

A key property of our learning dynamics is that the beliefs on opponent strategy and $Q$-functions are updated simultaneously though the update of the latter is at a slower timescale than the former. This is consistent with the literature on evolutionary game theory, e.g., see \cite{ref:Ely01,ref:Sandholm01} and also the closely related recent paper \cite{ref:Leslie20}, which we discuss below.

We show that beliefs on the opponent strategy converge to a (stationary) Nash equilibrium of zero-sum stochastic games under the assumption that each state is visited infinitely often in both the model-based case and model-free case. In the model-free case, players do not know their own payoff functions and the underlying state transition probabilities, however, each player can still observe her  realized stage payoff and the current state of the game, as in the reinforcement learning literature. Similarly, beliefs on $Q$-functions converge to the $Q$-functions associated with the equilibrium strategies in both cases. 

The fictitious play dynamics presented here reduce to the classical fictitious play when there is only one state and inherit the following features of the classical fictitious play: $i)$ The dynamics do not require knowledge of the underlying game's type and is not specific to any specific class of games. $ii)$ Following this scheme, players attain the best performance against an opponent following an asymptotically stationary strategy.
$iii)$ If the dynamics converge, it must converge to an equilibrium of the underlying game.

\subsubsection{Challenges}

As a first challenge, we note that even though the players play best response strategies in the auxiliary games associated with each state, the players do not necessarily play the same auxiliary game repeatedly because their payoff matrices, i.e., $Q$-functions, are time-varying and depend on the play at other states. Since these auxiliary game are not time-invariant, the convergence results for fictitious play or best response dynamics in zero-sum strategic-form games with repeated play, e.g., \cite{ref:Robinson51,ref:Harris98}, are not directly applicable to zero-sum stochastic games. We remove this dependency by approximating the discrete-time update via a differential inclusion (specific to each state) at the timescale of the beliefs on strategies as if the beliefs on $Q$-functions are time-invariant. To this end, we interpret an appropriate affine interpolation of the discrete-time update as a perturbed solution to a certain differential inclusion and characterize its limit set via a Lyapunov function argument, as shown by \cite{ref:Benaim05}. 

As a second challenge, we note that when the players do not share a common belief about the $Q$-function, their individual beliefs do not necessarily sum to zero due to the independent updates. Hence, the auxiliary games are non-zero-sum in general. This is problematic since it is well-known that fictitious play (or any uncoupled learning dynamic that does not incorporate the opponent's objective) cannot converge to an equilibrium in every class of non-zero-sum games \cite{ref:Hart03}. To address this challenge, we exploit the structure of the stochastic game and construct a new Lyapunov function for both zero-sum and non-zero-sum games. 

For zero-sum games, our Lyapunov function reduces to the Lyapunov function presented by \cite{ref:Harris98} for continuous-time best response dynamics in zero-sum strategic-form games. For non-zero sum games, this new Lyapunov function enables us to characterize the limit set of the beliefs on strategies in terms of how much the sum of beliefs on $Q$-functions deviates from zero. We exploit this characterization via the asynchronous stochastic approximation methods, provided by \cite{ref:Tsitsiklis94}, to show that the {\em beliefs} on $Q$-functions sum to zero asymptotically (although they do not necessarily sum to zero in finite time). We emphasize that the zero-sum structure of stage-payoffs of the underlying stochastic game is crucial for this result to hold.

\subsection{Related Works}

Our paper is most closely related to the recent paper \cite{ref:Leslie20} which presents and studies a continuous-time best-response dynamics for zero-sum stochastic games. They consider dynamics where each player selects {\em a mixed strategy in an auxiliary game} and updates her  strategy in the direction of her  best response to opponent's current mixed strategy in the auxiliary game. A single continuation payoff (common among the players) is updated at a slower speed representing the time average of the auxiliary game payoffs up to time $t$. The common update ensures that the auxiliary game is always zero-sum. This allows building on the convergence analysis provided by \cite{ref:Harris98} since two-timescale learning enables the mixed strategies to track an equilibrium associated with the estimates of the continuation-payoffs. In \cite{ref:Leslie20}, the authors generalized the convergence result in \cite{ref:Vigeral10} (that extends the convergence result in \cite{ref:Shapley53} to continuous-time dynamics) to settings with asymptotically negligible (tracking) error, thus establishing convergence of their dynamics in zero-sum stochastic games. 

Their dynamics involve updating mixed strategies at every state at every time. Hence, the authors study also an alternative update rule by considering a continuous-time embedding of the actual play of the stochastic game where game transitions according to a controlled continuous-time Markov chain. Our paper instead considers dynamics where each player follows a best response pure action in the auxiliary game (without any specific tie-breaking rule) while updating her  $Q$-function using her  belief on the opponent strategy and her  current $Q$-function estimate. Therefore, estimates of $Q$-functions do not necessarily sum to zero leading to an auxiliary game that is not necessarily zero-sum. Furthermore players update their beliefs on opponent strategy only for the current state within the course of stochastic game without need for such a continuous-time embedding.

Other related papers include \cite{ref:Shapley53}, \cite{ref:Vrieze82}, and \cite{ref:Schoenmakers07}. In \cite{ref:Shapley53}, the author presented and studied the minimax-value iteration in zero-sum stochastic games, which can be viewed as a generalization of value iteration in Markov decision problems to zero-sum stochastic game settings by replacing the optimization with the minimax-value of the auxiliary zero-sum game. \cite{ref:Shapley53} showed that the minimax-value iteration converges to a unique point, establishing existence of a stationary equilibrium in zero-sum stochastic games. The minimax value iteration necessitates computation of the minimax-value of the auxiliary game at each stage. This can be done by solving a collection of linear programs (LPs), one per state at each iteration which can  be computationally demanding. 

To mitigate the need to solve LPs at each iteration, in \cite{ref:Vrieze82}, the authors presented and studied a fictitious play like discrete-time update rule to find an equilibrium in zero-sum stochastic games without solving LPs. This update rule involves each player playing a best-response in an auxiliary game with a {\em common} continuation payoff as in \cite{ref:Leslie20}, which preserves the zero-sum nature of the auxiliary game. However, unlike \cite{ref:Leslie20}, in \cite{ref:Vrieze82}, the players update the continuation payoff using the payoff estimate of one of the players, which simplifies the analysis, but is not a natural update process. The dynamics reduce to fictitious play applied to a convergent sequence of zero-sum strategic-form games for each state. 

For {\em time-averaged} stochastic games,  in \cite{ref:Schoenmakers07}, the authors focused on a special class with two players, two states and two actions (per state), and provided an example in which the fictitious play presented does not necessarily converge to a stationary equilibrium. This is in contrast with results for two-player two-action strategic-form games, where fictitious play is known to converge to an equilibrium with a certain tie-breaking rule (see \cite{ref:Miyasawa61}), or when the game has the ``diagonal property" for any tie-breaking rule (see \cite{ref:Monderer96a,ref:Monderer96b}).

\subsubsection{Model-free Case}

Our paper is also related to a number of papers on multi-agent reinforcement learning, e.g., see the survey in \cite{ref:Zhang19} and the references therein. Particularly noteworthy is  \cite{ref:Littman94} which presented a model-free version of \cite{ref:Shapley53}'s minimax-value iteration via a Q-learning-type algorithm, called Minimax-Q. Similar to Shapley's minimax value iteration, Minimax-Q assumes a zero-sum structure and therefore is specific to zero-sum games. 

Alternative to Minimax-Q, in \cite{ref:Tesauro03}, the author presented a fictitious-play-like dynamics, called Hyper-Q, which applies beyond zero-sum games. Hyper-Q has dynamics similar to ours, however, evolves over a single timescale without any convergence guarantee in any specific class of stochastic games. On the other hand, in \cite{ref:Borkar02}, the author presented an actor-critic-type learning algorithm that is also not specific to zero-sum games. Contrary to Hyper-Q, there players do not seek to learn opponent strategies based on actions taken. He showed that a certain (weighted) empirical distribution of the joint actions taken converges to the set of (a modified version of) {\em generalized Nash equilibria} in stochastic games provided that each state-action pair is visited infinitely often and frequently enough. This is a weaker sense of convergence compared to our result although it is for stochastic games beyond zero-sum.

Other than the papers reviewed above, there are also several other multi-player reinforcement learning algorithms that are shown to have good convergence properties in stochastic games with respect to certain performance measures provided that every player follows rules which at times may not align with their best interests. For example,  in \cite{ref:Brafman02} and more recently \cite{ref:Bai20}, the authors focused on scenarios where players play in a coordinated manner a {\em finite-horizon-version} of a zero-sum stochastic game within repeated episodes, referred to as {\em episodic reinforcement learning}, even though player payoffs are defined over infinite horizon. In another line of work, in \cite{ref:Wei17} and \cite{ref:Arslan17}, the authors presented and studied algorithms that update policies only at certain time instances while keeping them {\em fixed} in between --even when players may have incentive to change their actions-- in order to create a stationary environment for learning the underlying model or estimating the associated $Q$-functions.

\subsection{Organization}
The rest of the paper is organized as follows. In \Cref{sec:model,sec:fictitiousplay}, we model stochastic games and our fictitious play scheme, respectively. We present the assumptions and the convergence results in \Cref{sec:main}. \arXiv{}{We provide preliminary information on a convergence result on asynchronous discrete-time iterations that we use in the proof of the convergence results in \Cref{app:aux}.} The proofs of the main convergence results in the model-based and model-free settings are provided, respectively, in \Cref{sec:proof} and \Cref{app:cor:main}. In \Cref{sec:example}, we provide an illustrative example. We conclude the paper with some remarks in \Cref{sec:conclusion}. 

\section{Stochastic Games}
\label{sec:model}

Consider two players that interact with each other by taking actions in a dynamic environment over an infinite horizon with discrete time $k=0,1,\ldots$. The players collect a \textit{stage payoff} depending on their actions and the current state of the environment, which also determines the next state. A two-player zero-sum stochastic game is a tuple $\langle  S , A,r,p, \gamma \rangle$ constructed as follows.
\begin{itemize}
\item Let $S$ be a set of {\em finitely} many states.
\item Let $A^i$ be the set of {\em finitely} many actions that \player{i} can take at any state $s\in S$.\footnote{This can be generalized to the case where action spaces depend on state straightforwardly.} Furthermore, $A:=A^1\times A^2$ denotes the set of action profiles $a=(a^1,a^2)$ for $a^i\in A^i$, $i=1,2$.
\item Let $r^{i}: S\times A \rightarrow \mathbb{R}$ denote the \textit{stage payoff function} of \player{i} at state $s$. Since it is a zero-sum game, we have $r^{1}(s,a) + r^2(s,a)=0$ for all $(s,a)\in S\times A$.
\item For any pair of states $(s,\tilde{s})$ and action profile $a\in A$, we define $p(\tilde{s}|s,a)$ as the {\em transition probability} from $s$ to $\tilde{s}$ given action profile $a$.
\item Let $\gamma \in(0,1)$ denote a \textit{discount factor} that affects the importance of future stage payoffs.
\end{itemize}

We focus on \textit{stationary (Markov) strategies}, meaning that at each stage each player plays a mixed action that depends only on the current state (and not for example on time). This does not cause any loss of generality due to the existence result in \cite{ref:Shapley53}. More specifically, for each $i=1,2$, we denote by $\pi^i(s,a^i)\in[0,1]$  the probability that \player{i} takes action $a^i$ at state $s$ and the stationary strategy of \player{i} by $\pi^i$. Let us also denote the strategy profile by $\pi:=\{\pi^1,\pi^2\}$. Correspondingly, $a_k= (a_k^1,a_k^2)$ is the action profile at stage $k$.  

We define the \textit{expected utility} of player $i$ under the strategy profile $\pi$ as the expected discounted sum of stage payoffs
\begin{align}\label{eq:utility}
U^i(\pi^1,\pi^2) := \mathbb{E}\left\{\sum_{k=0}^{\infty}\gamma^k r^{i}(s_k,a_{k})\right\},
\end{align}
where $\{s_k\sim p(\cdot|s_{k-1},a_{k-1})\}_{k>0}$ and $\{a_k\sim\pi(s_k,\cdot)\}_{k\geq 0}$ are stochastic processes, respectively, representing the state and the action profile at each stage $k$ and the expectation is taken with respect to all randomness induced by the initial state distribution $s_0\sim p_o\in \Delta(S)$, the state transition kernel and strategy profile $\pi$.\footnote{For a set $X$, we denote the probability simplex by $\Delta(X)$.}

A strategy profile $(\tilde{\pi}^1,\tilde{\pi}^2)$ is an $\varepsilon$-Nash equilibrium of the stochastic game with $\varepsilon\geq 0$ provided that
\begin{subequations}\label{eq:equilibrium}
\begin{align}
&U^1(\tilde{\pi}^1,\tilde{\pi}^2)\ge U^1(\pi^1,\tilde{\pi}^2) - \varepsilon \quad \textup{for all} \quad \pi^1, \\ 
&U^2(\tilde{\pi}^1,\tilde{\pi}^2)\ge U^2(\tilde{\pi}^1,\pi^2) - \varepsilon \quad \textup{for all} \quad \pi^2.
\end{align}
\end{subequations} 
Correspondingly, $(\tilde{\pi}^1,\tilde{\pi}^2)$ is a Nash equilibrium if \eqref{eq:equilibrium} holds with $\varepsilon = 0$.

\subsection{Auxiliary Stage-games in a Stochastic Game} At each stage of a stochastic game, the action profile determines the current stage-payoff and the stage-payoffs that will be received in future stages by determining the next state (since stage-payoffs also depend on the state). Correspondingly, if \player{i} knew that the opponent $-i$ is playing according to the stationary strategy $\pi^{-i}$, then the value of the action profile $a\in A$ at current state $s$, denoted by $Q^i(s,a)$ (and known as \textit{$Q$-function}), would satisfy the following fixed-point equation
\begin{equation}\label{eq:q}
Q^i(s,a) = r^i(s,a) + \gamma\sum_{\tilde{s}\in S} p(\tilde{s}|s,a) \max_{\tilde{a}^i\in A^i} \mathbb{E}_{\tilde{a}^{-i}\sim \pi^{-i}(\tilde{s},\cdot)}\{Q^i(\tilde{s},\tilde{a})\}.
\end{equation}
This follows from backward induction based on the principle that \player{i} would look for maximizing her expected utility, as described in \eqref{eq:utility}, in future stages. Therefore, a stochastic game can be viewed as a collection of \textit{auxiliary stage-games} specific to each state and represented by $\langle A^1,A^2,Q^1(s,\cdot),Q^2(s,\cdot)\rangle$. For notational convenience, we also define the value function $v^i:S\rightarrow \mathbb{R}$ by
\begin{equation}\label{eq:vis}
v^i(s) = \max_{a^i\in A^i} \mathbb{E}_{a^{-i}\sim \pi^{-i}(s,\cdot)}\{Q^i(s,a)\}
\end{equation}
which corresponds to the maximum value \player{i} would get in the associated auxiliary stage-game. Note that the dependence of $Q^i$ and $v^i$ on $\pi^{-i}$ is implicit in \eqref{eq:q} and \eqref{eq:vis} for notational convenience.

We also note that in two-player zero-sum stochastic games, there may exist multiple stationary equilibria in two-player zero-sum stochastic games. However, the $Q$-functions and value functions associated with any stationary equilibrium  are all the same \cite{ref:Shapley53}. We denote them, respectively, by $(Q_*^1,Q^2_*)$ and $(v_*^1,v_*^2)$.

Though a stochastic game can be viewed as a collection of such auxiliary stage-games that are being played repeatedly and asynchronously, the opponent's strategy and the $Q$-function are not readily available to the players. Furthermore, these auxiliary stage-games are not necessarily stationary. However, as in the classical fictitious play, the players can form beliefs on them based on the empirical play as if they are stationary. Given this observation, in the following section, we introduce fictitious-play-type learning dynamics that combines the classical fictitious play with the $Q$-learning for stochastic games.

\section{Fictitious Play in Stochastic Games}\label{sec:fictitiousplay}

We consider scenarios where players follow fictitious play dynamics in which they not only form beliefs on the opponent's strategy but also on the $Q$-function based on the history of the play. They take the greedy best action in the associated auxiliary stage-game conditioned on their beliefs. We emphasize that the players do not know the opponent's objective. In other words, they do not possess the knowledge that the underlying game is zero-sum.

In the following, we describe the learning dynamics for the typical \player{1} in both model-based and model-free settings. The dynamics for the typical opponent \player{2} is a mirror of it.

\subsection{Fictitious Play for the Model-based Setting}\label{sec:fictitious}

At each stage, \player{1} has beliefs on the opponent strategy and her $Q$-function, respectively, denoted by $\hat{\pi}_{k}^2:S\times A \rightarrow [0,1]$ and $\hat{Q}_{k}^1:S\times A\rightarrow \mathbb{R}$. For notational convenience, we define $\hat{\pi}_k^2(s) := \hat{\pi}_k^2(s,\cdot)$ and $\hat{Q}_k^1(s):=\hat{Q}_k^1(s,\cdot)$. At stage $k=0$, she initializes her  beliefs {\em arbitrarily} such that $\hat{\pi}_{0}^2(s)\in\Delta(A^2)$ and $\hat{Q}_{0}^1(s)\in \mathbb{R}^{|A^1|\times |A^2|}$ for each $s\in S$. 

Let $s\in S $ denote the current state at stage $k\geq 0$. \Player{1} and simultaneously \player{2} take their greedy best response actions $a_{k}^1\in A^1$ and $a_{k}^2\in A^2$. For example, \player{1} can take any action satisfying
\begin{align}\label{eq:maxx}
a_{k}^1 \in \argmax_{a^1\in A^1}\; \mathbb{E}_{a^2\sim \hat{\pi}_{k}^2(s)}\left\{\hat{Q}_{k}^1(s,a^1,a^2)\right\},
\end{align}
according to arbitrary tie-breaking rules. Without loss of generality, we consider pure actions as degenerate mixed strategies giving probability one to the associated action, i.e., $A^i\subset\Delta(A^i)$. The players can observe the opponent's action. Hence, \player{1} updates her belief on \player{2}'s strategy at the current state $s$ according to
\begin{equation}\label{eq:policy}
\hat{\pi}_{k+1}^2(s) = \hat{\pi}_{k}^2(s) + \alpha_{\#s} (a_{k}^2 - \hat{\pi}_{k}^2(s)),
\end{equation}
where $\alpha_{\#s}\in(0,1]$ is a step-size specific to $\#s$, representing the number of times that $s$ gets visited until (and including) stage $k$. 

Furthermore, \player{1} updates her  belief on her  own $Q$-function only for the current state $s$. The update of $\hat{Q}_{k}^1(s)$ is given by
\begin{align}
\hat{Q}_{k+1}^1(s,a) = \hat{Q}_{k}^1(s,a) + \beta_{\#s}\left(r^1(s,a) + \gamma \sum_{\tilde{s}\in S}p(\tilde{s}|s,a)\hat{v}_{k}^1(\tilde{s})-\hat{Q}_{k}^1(s,a)\right),\label{eq:Qfunc}
\end{align}
for all $a\in A$, where we define the value function estimate $\hat{v}_{k}^1:S\rightarrow\mathbb{R}$ by 
\begin{equation}\label{eq:vbv}
\hat{v}_{k}^1(\tilde{s}) := \max_{\tilde{a}^1\in A^1} \mathbb{E}_{\tilde{a}^{2}\sim\hat{\pi}_k^{2}(\tilde{s})} \left\{\hat{Q}_{k}^1(\tilde{s},\tilde{a})\right\}
\end{equation}
and again $\beta_{\#s}\in(0,1]$ is a step-size specific to $\#s$. 

\begin{table}[t]
\caption{Fictitious play of the typical \player{i} in stochastic games.}
\label{algo1}
\hrule
\begin{algorithmic}[1]
\Require Keep track of $\hat{\pi}_k^{-i}$ and $\hat{Q}_k^i$ for every $(s,a)$.
\For{Each stage $k\geq 0$}
\State {\bf Observe} the current state $s_k$.
\State {\bf Take action} $a^i_k$ according to \eqref{eq:maxx}.
\State {\bf Observe} the opponent's action $a^{-i}_k$.
\State {\bf Update} $\hat{\pi}_{k}^{-i}(s_k)$ according to \eqref{eq:policy}.
\State {\bf Update} $\hat{Q}_{k}^i(s_k,a)$ for all $a\in A$ according to \eqref{eq:Qfunc}.
\EndFor
\end{algorithmic} 
\hrule
\end{table}

\Player{1} does not update her beliefs associated with other states, i.e., $\hat{\pi}_{k+1}^2(s')=\hat{\pi}_k^2(s')$ and $\hat{Q}_{k+1}^1(s')=\hat{Q}_k^1(s')$ if $s'\neq s$, i.e., if $s'$ is not the current state. A description of the dynamics is tabulated in Table \ref{algo1}.

Note that given the definition of $\hat{v}_{k}^i(s)$ in \eqref{eq:vbv}, we do not necessarily have $\hat{v}_{k}^1(s)+\hat{v}_{k}^2(s)=0$ for all $s\in S $. Moreover, when $\hat{v}_{k}^1(s)+\hat{v}_{k}^2(s)\neq0$ for some $s\in S $, then by \eqref{eq:Qfunc}, we do not necessarily have $\hat{Q}_{k+1}^1(s)+\hat{Q}_{k+1}^2(s)$ equal to the zero matrix for all $s\in S $. Therefore, the auxiliary stage-games need not be zero-sum in this learning dynamics. 

Alternatively, consider the scenario where \player{1} and \player{2} update their beliefs on $Q$-functions as in \eqref{eq:Qfunc} but with
\begin{equation}\label{eq:vbv2}
\tilde{v}_{k}^i(\tilde{s}) = \mathbb{E}_{\tilde{a}\sim \hat{\pi}(\tilde{s})} \left\{\hat{Q}_{k}^i(\tilde{s},\tilde{a})\right\}.
\end{equation}
In other words, the players form beliefs on their own strategies (which is not a natural dynamics but we pursue it briefly to illustrate the more tractable mathematical structure this leads to). If the beliefs on $Q$-functions are initialized such that $\hat{Q}_{0}^1(s)+\hat{Q}_{0}^2(s)$ is equal to the zero matrix for each $s$, then we have $\tilde{v}_{0}^1(s)+\tilde{v}_{0}^2(s)=0$. Then by induction, it can be shown that $\tilde{v}_{k}^1(s)+\tilde{v}_{k}^2(s)=0$ for all $s$ and $k$, and indeed $\hat{Q}_{k}^1(s)+\hat{Q}_{k}^2(s)$ is equal to the zero matrix for all $s$ and $k$. Therefore, the auxiliary stage-games would always be zero-sum.

In the scenarios where the auxiliary stage-games remain always zero-sum, the convergence analysis is a direct application of the two-timescale stochastic approximation theory built on the convergence result for fictitious play in zero-sum strategic-form games (with repeated play) provided by \cite{ref:Harris98} and the convergence result for the minimax value iteration provided by \cite{ref:Shapley53}. However, this is not the case when players follow an uncoupled learning dynamics, such as our two-timescale fictitious play, and its convergence analysis necessitates development of new technical tools specific to the structure of stochastic games rather than resorting directly to the two-timescale stochastic approximation theory.

\subsection{Fictitious Play for the Model-free Setting}\label{sec:modelFree}

Next we consider the scenarios where players do not know their own stage payoff function and the transition probabilities. They can still observe their current stage payoff (realized), current state (visited), and the current action (taken by the opponent). Given the beliefs on $Q$-functions, the players take the actions according to \eqref{eq:maxx} while they may also take some random action with some small probability $\epsilon>0$ to {\em experiment} stochastic state transitions, as in \cite{ref:Littman94}. For example, \player{1} can take action
\begin{equation}\label{eq:explore}
a_k^1 = \left\{\begin{array}{ll} a_*^1 &\mathrm{w.p. } \;(1-\epsilon)\\
u^1 &\mathrm{w.p. }\;\epsilon\end{array}\right.
\end{equation}
where $a_*^1$ is a greedy best response satisfying \eqref{eq:maxx} while $u^1\sim \mathcal{U}(A^1)$ with $\mathcal{U}(\cdot)$ denoting the uniform distribution over the associated set.
 We focus on this basic exploration strategy as a proof of concept. The players can also resort to more sophisticated strategies to speed up their exploration, e.g., see \cite{ref:Lattimore20}.  

Players still update their beliefs on opponent strategy according to \eqref{eq:policy}. However since \player{1} and \player{2} cannot update their beliefs on $Q$-functions as in \eqref{eq:Qfunc} without knowing state transition probabilities, they instead follow a $Q$-learning-type of update described as follows.

\begin{table}[t]
\caption{Model-free fictitious play of the typical \player{i} in stochastic games.}
\label{algo2}
\hrule
\begin{algorithmic}[1]
\Require Keep track of $\hat{\pi}_k^{-i}$ and $\hat{Q}_k^i$ for every $(s,a)$.
\For{Each stage $k\geq 0$}
\State {\bf Observe} the current state $s_k$.
\State {\bf Update} $\hat{Q}_{k-1}^i(s_{k-1},a_{k-1})$ according to \eqref{eq:QfuncFree}.
\State {\bf Take action} $a^i_k$ according to \eqref{eq:explore}.
\State {\bf Observe} the opponent's action $a^{-i}_k$.
\State {\bf Update} $\hat{\pi}_{k}^{-i}(s_k)$ according to \eqref{eq:policy}.
\EndFor
\end{algorithmic} 
\hrule
\end{table}

Player $i$ observes current state $s$, current action profile $a$, and her current stage payoff (denoted by $r_{k}^i$), and by looking one-stage ahead, she also observes the next state $\tilde{s}$. Given the triple $(s,a,\tilde{s})$, she uses an estimate for the continuation payoff for the next state, i.e., $\hat{v}_{k}^i(\tilde{s})$, as an unbiased estimator of $\sum_{s'\in S }\hat{v}_{k}^i(s')p(s'|s,a)$. She updates her belief on the $Q$-function only for the current state and action profile $(s,a)$, according to 
\begin{align}
\hat{Q}_{k+1}^i(s,a) = \hat{Q}_{k}^i(s,a) + \beta_{\#(s,a)} \left(r_{k}^i + \gamma \hat{v}_{k}^i(\tilde{s}) - \hat{Q}_{k}^i(s,a)\right),\label{eq:QfuncFree}
\end{align}
where $\hat{v}_{k}^i$ is as described in \eqref{eq:vbv}. Note that here $\beta_{\#(s,a)}\in(0,1]$ is a step-size specific to $\#(s,a)$ representing the number of times \textit{state-action} pair $(s,a)$ occurs until (and including) stage $k$. Note also that \player{1} does not update $\hat{Q}_k^1$ associated with other state-action pairs, i.e., $\hat{Q}_{k+1}^1(s',a')=\hat{Q}_k^1(s',a')$ if $(s',a')\neq (s,a)$, i.e., if either $s'$ is not the current state or $a'$ is not the current action profile. A description of the model-free dynamics is tabulated in Table \ref{algo2}. Note that $\hat{Q}_{k-1}^i(s_{k-1},a_{k-1})$ gets updated after $s_k$ is observed. Therefore, the updates of beliefs take place at different orders in Tables \ref{algo1} and \ref{algo2}.

\section{Main Result}\label{sec:main}

In this paper, we focus on whether the beliefs formed on the opponent's strategies and $Q$-functions converge to a stationary equilibrium and the corresponding $Q$-functions in zero-sum stochastic games, or not. The answer is {\em affirmative} for both model-based and model-free settings under certain assumptions provided below precisely. 

\begin{assumption}\label{assume:visits}
Each state is visited infinitely often with probability one.
\end{assumption}

Players update their beliefs associated with a state only when that state is visited. This assumption ensures that players have sufficient time to revise and improve their beliefs. Furthermore, it holds, if the stochastic game is {\em irreducible}, e.g., transition probabilities between any pair of states are positive for any joint action as in \cite{ref:Leslie20}.    

\begin{assumption}\label{assume:steps}
The step sizes $\{\alpha_c\in(0,1]\}_{c= 0}^{\infty}$ and $\{\beta_c\in(0,1]\}_{c= 0}^{\infty}$ satisfy
$\sum_{c=0}^{\infty}\alpha_{c} = \infty,\, \sum_{c=0}^{\infty}\beta_{c} = \infty$,  and $\lim_{c\rightarrow \infty}\alpha_c = \lim_{c\rightarrow\infty} \beta_c = 0$. The beliefs on $Q$-functions are updated at a slower timescale compared to the timescale in which the beliefs on strategies are updated, i.e., 
$
\lim_{c\rightarrow\infty} \frac{\beta_{c}}{\alpha_{c}} = 0.
$
\end{assumption}

Now we are ready to present the convergence results specific to zero-sum stochastic games.

\begin{theorem}\label{thm:main}
Suppose that \Cref{assume:visits,assume:steps} hold. When both players follow the fictitious play dynamics described in Table \ref{algo1}, i.e., \eqref{eq:maxx}-\eqref{eq:Qfunc}, the beliefs on strategies and $Q$-functions, respectively, converge to a stationary equilibrium and the corresponding $Q$-functions in zero-sum stochastic games almost surely. In other words, for some stationary equilibrium $\pi_* = (\pi^1_*,\pi^2_*)$, we have 
$(\hat{\pi}_{k}^1,\hat{\pi}_{k}^2) \rightarrow (\pi^{1}_*,\pi^{2}_*)$ and $(\hat{Q}_{k}^1,\hat{Q}_{k}^2) \rightarrow (Q^{1}_{*},Q^{2}_{*})$,
as $k\rightarrow \infty$, with probability $1$.
\end{theorem}

The following corollary to \Cref{thm:main} characterizes the convergence properties of the dynamics in two-player \textit{general-sum} stochastic games in terms of how much the stage-payoffs deviate from the zero-sum structure. Particularly, it shows convergence of the dynamics to a \textit{near} equilibrium in \textit{near} zero-sum stochastic games.

\begin{corollary}\label{cor:dev}
Suppose that \Cref{assume:visits,assume:steps} hold and both players follow the fictitious play dynamics described in Table \ref{algo1}, i.e., \eqref{eq:maxx}-\eqref{eq:Qfunc}. Then, in two-player general-sum stochastic games, we have
\begin{equation}
\limsup_{k\rightarrow\infty} |\hat{Q}_k^i(s,a) - Q_d^i(s,a)| \leq \frac{d(1+\gamma)}{\gamma(1-\gamma)^2},\quad\forall (s,a),
\end{equation}
with probability $1$, where $Q_d^i:S\times A \rightarrow \mathbb{R}$ satisfies the following fixed point equation
\begin{equation}
Q_d^i(s,a) = r^i(s,a) + \gamma\sum_{\tilde{s}\in S} p(\tilde{s}|s,a) \max_{\tilde{\pi}^i\in \Delta(A^i)}\min_{\tilde{\pi}^{-i}\sim\Delta(A^{-i})} \mathbb{E}_{(\tilde{a}^i,\tilde{a}^{-i})\sim (\tilde{\pi}^i,\tilde{\pi}^{-i})}\{Q_d^i(\tilde{s},\tilde{a})\},\nonumber
\end{equation}
for all $(s,a)$ and $d := \max_{(s,a)}|r^1(s,a)+r^2(s,a)|$.
\end{corollary}

In the model-free setting, players can update only a single entry of the belief on $Q$-function corresponding to the current action profile. This strengthens the coupling across states and makes the dynamics difficult to track. Furthermore, the players can only observe the realization of the state transitions, which introduces stochastic approximation errors in the learning dynamics. Hence, we make the following assumption to limit the impact of the coupling and the stochastic approximation errors.

\begin{assumption}\label{assume:stepFree}
 The step sizes satisfy $\sum_{c=0}^{\infty}\alpha_c^2 <\infty$ and $\sum_{c=0}^{\infty}\beta_c^2 <\infty$. The sequence $\{\beta_c\}_{c\geq 0}$ is monotonically decreasing. We have
 $\lim_{c\rightarrow\infty} \frac{\beta_{\lfloor mc\rfloor}}{\alpha_c} = 0$ for any $m\in(0,1]$.
\end{assumption}

The first part of \Cref{assume:stepFree} ensures that the stochastic approximation terms are square integrable Martingale difference sequences conditioned on the history. On the other hand, the second part of \Cref{assume:stepFree} ensures that $\beta_{\#(s,a)}/\alpha_{\#s}$ gets arbitrarily small asymptotically even though $\#(s,a)$ increases more slowly than $\#s$. We also emphasize that \Cref{assume:steps,assume:stepFree} are properties of step sizes and do not impose further conditions on zero-sum stochastic games for which the results hold. For example, the step sizes given by 
$\alpha_c = (c+1)^{-\rhoa}$ and $\beta_c = (c+1)^{-\rhob}$, where $1/2 < \rhoa < \rhob \leq 1$, satisfy \Cref{assume:steps,assume:stepFree}.

The following theorem characterizes the convergence properties of fictitious play for the model-free setting.

\begin{theorem}\label{cor:main}
Suppose that \Cref{assume:visits,assume:steps,assume:stepFree} hold. When both players follow the fictitious play dynamics described in Table \ref{algo2}, i.e., \eqref{eq:explore}, \eqref{eq:policy}, and \eqref{eq:QfuncFree}, the beliefs on strategies and $Q$-functions converge to a near equilibrium and the equilibrium $Q$-functions with an approximation level linear in the exploration probability $\epsilon>0$, almost surely. More explicitly, we have
\begin{align}
&\limsup_{k\rightarrow\infty} \left(U^i(\pi^i,\hat{\pi}_k^{-i}) - U^i(\hat{\pi}_k^i,\hat{\pi}_k^{-i})\right) \leq 2 \epsilon D \, \frac{(1+\gamma)^2}{\gamma(1-\gamma)^3},\quad \forall \pi^i,\\
&\limsup_{k\rightarrow\infty} |\hat{Q}_{k}^i(s,a) - Q_{*}^i(s,a)| \leq \epsilon D \, \frac{1+\gamma}{(1-\gamma)^2},\quad \forall (s,a)\label{eq:Qresult}
\end{align}
with probability $1$, where $D= \frac{1}{1-\gamma}\sum_{i} \max_{(s,a)} |r^i(s,a)|$.
\end{theorem}

Note that if the players decrease their exploration probability at a suitable rate, the learning dynamics for the model-free case can also converge to an exact equilibrium of the stochastic game, e.g., see \cite[Section 5]{ref:Leslie06}. Note also that the analysis can be generalized to the case with independent random perturbations of stage-payoffs (with compact support) straightforwardly, as shown in the extended version \cite[Section 6.3]{ref:Sayin20}.

We provide the proofs of \Cref{thm:main,cor:main} in \Cref{sec:proof,app:cor:main}, respectively. 

\arXiv{}{
In the following section, we provide a theorem characterizing the convergence properties of asynchronous discrete-time iterations to be used in the proofs of  \Cref{thm:main,cor:main}.

\section{On the Convergence of Asynchronous Discrete-time Iterations}\label{app:aux}

The following theorem follows from a slight modification of the result in \cite[Theorem 3]{ref:Tsitsiklis94} to address the impact of  error terms that are asymptotically bounded (or negligible as a special case). 

\begin{theorem}\label{lem:aux}
Consider a sequence of vectors $\{y_k\}_{k= 0}^{\infty}$ such that the $n$th entry, denoted by $y_{k}[n]$, satisfies the following upper and lower bounds:
\begin{subequations}
\begin{align}
&y_{k+1}[n] \leq (1-\beta_{n,k}) y_{k}[n] + \beta_{n,k}(\gamma \|y_k\|_{\infty} +  \oepsilon_{k} + \omega_{n,k}),\label{eq:yy}\\
&y_{k+1}[n] \geq (1-\beta_{n,k}) y_{k}[n] + \beta_{n,k}(-\gamma \|y_k\|_{\infty} + \uepsilon_{k} + \omega_{n,k}),\label{eq:yy2}
\end{align}
\end{subequations}
where 
\begin{itemize}
\item a scalar discount factor $\gamma\in(0,1)$, 
\item the step size $\beta_{n,k}\in[0,1]$ satisfies $\sum_{k=0}^{\infty}\beta_{n,k} = \infty$, $\lim_{k\rightarrow\infty}\beta_{n,k} = 0$ for each $n$ with probability $1$,\footnote{The condition that $\sum_{k=0}^{\infty}\beta_{n,k} = \infty$ for each $n$ implies that the non-negative $\beta_{n,k}$ is positive infinitely many times as $k\rightarrow\infty$ (and therefore, each entry of the vector gets updated infinitely often).} 
\item the error terms satisfy 
\begin{equation}\label{eq:limsup}
\limsup_{k\rightarrow \infty} |\oepsilon_k|\leq c\quad \mbox{and}\quad\limsup_{k\rightarrow\infty}|\uepsilon_k|\leq c,
\end{equation}
with probability $1$,
\item the stochastic approximation term $\{\omega_{n,k}\}_{k=0}^{\infty}$ satisfies $\E{\omega_{n,k}|\mathfrak{F}_k} = 0$ and $\E{\omega_{n,k}^2|\mathfrak{F}_k}\leq C$ for each $n$ for some constant $C$ with respect to an increasing sequence of $\sigma$-fields $\mathfrak{F}_k:=\sigma(y_l,\{\omega_{n,l}\}_n,l\leq k)$.
\end{itemize} 
Suppose that $\|y_k\|_{\infty}\leq T$ for all $k$. Then we have 
\begin{equation}\label{eq:limsupY}
\limsup_{k\rightarrow \infty}\|y_{k}\|_{\infty} \leq \frac{c}{1-\gamma},
\end{equation}
with probability $1$, provided that either $\omega_{n,k} = 0$ for all $n,k$ or $\sum_{k=0}^{\infty} \beta_{n,k}^2 <\infty$ for each $n$ with probability $1$.
\end{theorem}


\begin{proof}
The proof follows from $i)$ constructing an auxiliary sequence $\{D^t\}_{t=0}^{\infty}$ that converges to zero over a separate time-scale $t=0,1,\ldots$ and then $ii)$ showing that there exists a sequence $\{k^t\}_{t\geq 0}$ such that $\|y_k\|_{\infty}\leq D^t+\frac{c}{1-\gamma}$ for all $k\geq k^t$.\footnote{We use the superscript to distinguigh the time-scale in use.}

{\em Step $i)$} Since $\gamma\in(0,1)$, fix some $\varepsilon>0$ such that $\gamma + 2\varepsilon<1$. By \eqref{eq:limsup} and $\|y_k\|_{\infty}\leq D$ for all $k$, there exists $k^0$ and $D^0$ such that
\begin{equation}\label{eq:defi}
\|y_k\|_{\infty} \leq D^0 + \frac{c}{1-\gamma},\;|\oepsilon_k| \leq \varepsilon D^0+c, \mbox{ and } |\uepsilon_k| \leq \varepsilon D^0 + c,
\end{equation}
for all $k\geq k^0$. We define a sequence $\{D^t\}_{t=0}^{\infty}$, initialized by $D^0$ and evolving according to
\begin{equation}\label{eq:DDD}
D^{t+1} = (\gamma+2\varepsilon)D^t,\;\forall t\geq0.
\end{equation}
Since $\gamma+2\varepsilon\in(0,1)$, we have $\lim_{t\rightarrow\infty} D^t = 0$. 

{\em Step $ii)$} Suppose that for some $t>0$, there exists $k^t$ such that the following holds:
\begin{equation}\label{eq:must}
\|y_k\|_{\infty} \leq D^t+\frac{c}{1-\gamma},\;|\oepsilon_k| \leq \varepsilon D^t+c, \mbox{ and } |\uepsilon_k| \leq \varepsilon D^t + c,
\end{equation}
for all $k\geq k^t$. Our goal is to show that there exists $k^{t+1}\geq k^t$ such that 
\begin{equation}\label{eq:goalY}
\|y_k\|_{\infty} \leq D^{t+1}+\frac{c}{1-\gamma},\;|\oepsilon_k| \leq \varepsilon D^{t+1}+c, \mbox{ and } |\uepsilon_k| \leq \varepsilon D^{t+1} + c,
\end{equation}
for all $k\geq k^{t+1}$. 

Let us start by restraining the error terms $\uepsilon_k,\oepsilon_k$ and $\omega_{n,k}$ via the auxiliary sequences $\{Y_{k}^t\}_{k=k^t}^{\infty}$ and $\{W_{k}^t\}_{k=k^t}^{\infty}$ defined by the following recursions:
\begin{equation}\label{eq:YYY}
Y_{k}^t[n] := \left\{\begin{array}{ll} 
D^t + \frac{c}{1-\gamma}&\mbox{if } k=k^t\\
(1-\beta_{n,k-1})Y_{k-1}^t[n] + \beta_{n,k-1}\left[(\gamma+\varepsilon) D^t+\frac{c}{1-\gamma}\right]&\mbox{if } k>k^t
\end{array}\right.
\end{equation}
and
\begin{equation}\label{eq:WWW}
W_{k}^t[n] := \left\{\begin{array}{ll} 
0 &\mbox{if } k=k^t\\
(1-\beta_{n,k-1})W_{k-1}^t[n] + \beta_{n,k-1}\omega_{n,k-1}&\mbox{if } k>k^t
\end{array}\right.,
\end{equation}
respectively. The following lemma highlights the role of these two auxiliary sequences. 

\begin{lemma}
By induction, we have
\begin{equation}\label{eq:Ybound}
-Y_{k}^t[n] + W_{k}^t[n] \leq y_{k}[n] \leq Y_{k}^t[n] + W_{k}^t[n],
\end{equation}
for all $k\geq k^t$.
\end{lemma}

\begin{proof}
By definitions \eqref{eq:YYY} and \eqref{eq:WWW}, we already have the inequality \eqref{eq:Ybound} for $k=k^t$. Now suppose that the inequality \eqref{eq:Ybound} holds for some $k>k^t$. Then by the upper bound on $y_{k+1}[n]$, as described in \eqref{eq:yy}, and the assumption \eqref{eq:must}, we have
\begin{align*} 
y_{k+1}[n] &\leq (1-\beta_{n,k}) y_{k}[n] + \beta_{n,k}(\gamma \|y_k\|_{\infty} +  \oepsilon_{k} + \omega_{n,k})\\
&\leq (1-\beta_{n,k}) (Y_{k}^t[n] + W_{k}^t[n]) + \beta_{n,k}\left[\gamma \left(D^t+\frac{c}{1-\gamma}\right) + \varepsilon D^t + c +\omega_{n,k}\right]\\
&= Y_{k+1}^t[n] + W_{k+1}^t[n]
\end{align*}
while by the lower bound on $y_{k+1}[n]$, as described in \eqref{eq:yy2}, and \eqref{eq:must}, we have
\begin{align*} 
y_{k+1}[n] &\geq (1-\beta_{n,k}) y_{k}[n] + \beta_{n,k}(-\gamma \|y_k\|_{\infty} +  \uepsilon_{k} + \omega_{n,k})\\
&\geq (1-\beta_{n,k}) (-Y_{k}^t[n] + W_{k}^t[n]) + \beta_{n,k}\left[-\gamma\left(D^t+\frac{c}{1-\gamma}\right) - (\varepsilon D^t + c) + \omega_{n,k}\right]\\
&= -Y_{k+1}^t[n] + W_{k+1}^t[n].
\end{align*}
Therefore we obtain \eqref{eq:Ybound}.
\end{proof}

The convergence properties of $Y_{k}^t[n]$ and $W_{k}^t[n]$ are well-known when $\{\beta_{n,k}\}$ satisfies certain conditions. For example the term $\left[(\gamma+\varepsilon)D^t+\frac{c}{1-\gamma}\right]$ in the recursion of $Y_{k}^t[n]$, \eqref{eq:YYY}, is fixed for all $k>k^t$. Therefore, $Y_{k}^t[n]$ evolves as a convex combination of the previous iterate and this fixed term, and gets closer and closer to the fixed term since $\sum_{k=0}^{\infty}\beta_{n,k}=\infty$. On the other hand $\{W_{k}^t[n]\}$ corresponds to the classical stochastic gradient algorithm for minimizing a quadratic cost function, whose convergence is well-established if the step sizes also satisfy $\sum_{k=0}^{\infty}\beta_{n,k}^2<\infty$, e.g., see \cite{ref:Poljak73}. Therefore we have
\begin{equation}
\lim_{k\rightarrow\infty} Y_{k}^t[n] = (\gamma+\varepsilon) D^t+\frac{c}{1-\gamma}, \mbox{ and }\lim_{k\rightarrow\infty} W_{k}^t[n] = 0,\;\forall n, \label{eq:Ylim}
\end{equation}
with probability $1$. Therefore, there exists $k^{t+1}\geq k^t$ such that $|Y_k^t[n]+W_k^t[n]|\leq (\gamma+2\varepsilon)D^t + \frac{c}{1-\gamma}$, almost surely, and $|\oepsilon_k|\leq \varepsilon D^{t+1}+c$ and $|\uepsilon_k|\leq \varepsilon D^{t+1}+c$. We obtain \eqref{eq:goalY} since $D^{t+1}=(\gamma+2\varepsilon)D^t$. By induction, \eqref{eq:must} holds for all $t$, and the fact that $D^t\rightarrow 0$ as $t\rightarrow \infty$ yields \eqref{eq:limsupY}.
\end{proof}
}

\section{Proof of \Cref{thm:main}: Convergence for the Model-based Setting}\label{sec:proof}

We divide the proof into three main steps: $i)$ We first decouple the dynamics across states over the fast timescale. $ii)$ We next zoom into the local dynamics specific to each state and characterize their limit set via a novel Lyapunov function argument addressing the deviation of auxiliary stage-games from zero-sum structure in two-player zero-sum stochastic games. $iii)$ We then zoom out to global dynamics across every state over the slow timescale and show that the beliefs on $Q$-functions sum to zero asymptotically, and then show that estimates of continuation payoffs track the minimax values associated with the beliefs on $Q$-functions. Finally, we show that the beliefs on $Q$-functions converge to the unique $Q$-functions of an equilibrium of the underlying zero-sum stochastic game.

We can now delve into the technical details.\footnote{We omit the qualification ``with probability $1$" since we have already discarded the suitable set of measure zero, where \Cref{assume:visits} does not hold.}

\subsection{Step $i)$ Decoupling the dynamics at the fast timescale} 

Let $s_k$ denote the current state at stage $k$.
Based on \eqref{eq:policy}, we can write the updates of the beliefs on strategies for state $s\in S $ as
\begin{align}\label{eq:pip}
\begin{bmatrix} \hat{\pi}_{k+1}^1(s) \\ \hat{\pi}_{k+1}^2(s) \end{bmatrix} = \begin{bmatrix} \hat{\pi}_{k}^1(s) \\ \hat{\pi}_{k}^2(s) \end{bmatrix} + \bar{\alpha}_{k}(s) \begin{bmatrix}a_{k}^{1}- \hat{\pi}_{k}^1(s)\\a_{k}^2- \hat{\pi}_{k}^2(s) \end{bmatrix}
\end{align}
where $\bar{\alpha}_{k}(s) := \indicator{s=s_k} \alpha_{\#s}\in[0,1]$.\footnote{We represent the indicator function (which is $1$ if $s=s_k$ and $0$ otherwise) by $\indicator{s=s_k}$.} At the same timescale, we can write the updates of the beliefs on $Q$-functions of both players for state $s\in S $ as
\begin{equation}\label{eq:Qeps}
\begin{bmatrix} \hat{Q}_{k+1}^1(s,a) \\ \hat{Q}_{k+1}^2(s,a) \end{bmatrix} = \begin{bmatrix} \hat{Q}_{k}^1(s,a) \\ \hat{Q}_{k}^2(s,a) \end{bmatrix} + \bar{\alpha}_{k}(s) \begin{bmatrix} \mathcal{E}_{k}^1(s,a) \\ \mathcal{E}_{k}^2(s,a) \end{bmatrix},\quad\forall a\in A,
\end{equation}
combining \eqref{eq:Qfunc} for each $i=1,2$ together. The error terms $\mathcal{E}_{k}^i:S\times A\rightarrow\mathbb{R}$ are defined by
\begin{align}\label{eq:err}
\begin{bmatrix} \mathcal{E}_{k}^1(s,a) \\ \mathcal{E}_{k}^2(s,a) \end{bmatrix} := \frac{\beta_{\#s}}{\alpha_{\#s}}\begin{bmatrix}r^1(s,a) + \gamma \sum_{\tilde{s}\in S } p(\tilde{s}|s,a)\hat{v}_{k}^1(\tilde{s}) - \hat{Q}_{k}^1(s,a)\\
r^2(s,a) + \gamma \sum_{\tilde{s}\in S }p(\tilde{s}|s,a)\hat{v}_{k}^2(\tilde{s})  - \hat{Q}_{k}^2(s,a)\end{bmatrix},
\end{align}
where $\hat{v}_{k}^i(\tilde{s})$ is as described in \eqref{eq:vbv}. Note that the iterates are bounded since the stage payoffs have compact support, the discount factor $\gamma \in (0,1)$, and step sizes take values only in $[0,1]$. Therefore, \Cref{assume:steps}, i.e., $\lim_{c\rightarrow\infty} \frac{\beta_c}{\alpha_c} = 0$, yields that the error matrices in \eqref{eq:Qeps} are asymptotically negligible.

Note that \eqref{eq:pip} and \eqref{eq:Qeps} for state $s$ are coupled with the dynamics at other states $s'\neq s$ through the asymptotically negligible error terms. We consider the evolution of the iterates specific to a state, $\x_k(s):=(\hat{\pi}_{k}^1(s),\hat{\pi}_{k}^2(s),\hat{Q}_{k}^1(s),\hat{Q}_{k}^2(s))$, {\em separately} by exploiting this weak coupling. Note that we have 
$\sum_{k=0}^{\infty}\bar{\alpha}_k(s) = \infty$
and $\bar{\alpha}_k(s)\rightarrow 0$ as $k\rightarrow\infty$ for each $s$ based on \Cref{assume:visits,assume:steps}. Therefore, the stochastic differential inclusion theory, e.g., see \cite[Theorem 3.6 and Proposition 3.27]{ref:Benaim05}, yields that we can characterize the limit set of $\x_k(s)$ by formulating a Lyapunov function for the following flow:\footnote{Note that $\mathbb{E}_{a^{2}\sim\hat{\pi}^{2}_k(s)}\{\hat{Q}_k^1(s,a)\} = (a^1)^T\hat{Q}_k^1(s)\hat{\pi}_k^2(s)$ and $\mathbb{E}_{a^{1}\sim\hat{\pi}^{1}_k(s)}\{\hat{Q}_k^2(s,a)\} = \hat{\pi}_k^1(s)^T\hat{Q}_k^2(s)a^2$.}
\begin{subequations}\label{eq:brd}
\begin{align}
&\frac{d\mypi^1(t)}{dt} + \mypi^1(t) \in \argmax_{a\in A^1} \left\{a^T\Q^1(t)\mypi^2(t)\right\} \label{eq:pidiff1}\\
&\frac{d\mypi^2(t)}{dt} + \mypi^2(t) \in \argmax_{a\in A^2} \left\{\mypi^1(t)^T\Q^2(t)a\right\} \label{eq:pidiff2}\\
&\frac{d\Q^i(t)}{dt} = O, \quad i=1,2\label{eq:Qdiff},
\end{align}
\end{subequations}
where we drop the dependence on $s$ for notational brevity, use a different notation for the functions $\mypi^i:[0,\infty)\rightarrow \Delta(A^i)$ and $\Q^i:[0,\infty)\rightarrow \mathbb{R}^{|A^1|\times|A^2|}$ for $i=1,2$, and represent zero matrices by $O$. Note that \eqref{eq:Qdiff} yields that $\Q^1(\cdot)$ and $\Q^2(\cdot)$ are time-invariant, e.g., $\tilde{Q}^1 := \Q^1(t)$ and $\tilde{Q}^2 := \Q^2(t)$ for all $t\in[0,\infty)$. Therefore, \eqref{eq:pidiff1} and \eqref{eq:pidiff2} correspond to the continuous-time best-response dynamics in a strategic-form game with payoff matrices $\tilde{Q}^1$ and $\tilde{Q}^2$. However, $\tilde{Q}^1+\tilde{Q}^2$ is not necessarily equal to the zero matrix. In other words, it is not necessarily a zero-sum game. In the next step, we formulate a novel Lyapunov function addressing this challenge.

\subsection{Step $ii)$ Zooming into the local dynamics at the fast timescale}

It is instructive to examine the continuous-time best response dynamics in zero-sum strategic-form games. For a {\em fixed} (absolutely continuous) solution to \eqref{eq:brd}, e.g., $(\mypi^1(t),\mypi^2(t))$, \cite[Section 4]{ref:Harris98} showed that
\begin{align}
V_H(\mypi^1(t),\mypi^2(t)) = \h^1(t) + \h^2(t),\label{eq:Harris}
\end{align}
where
\begin{equation}\label{eq:hij}
\h^1(t) := \max_{a\in A^1} \left\{a^T\tilde{Q}^{1} \mypi^{2}(t)\right\}\quad \mbox{and}\quad \h^2(t) := \max_{a\in A^2}  \left\{\mypi^1(t)^T \tilde{Q}^{2} a\right\}
\end{equation}
is a Lyapunov function for \eqref{eq:brd} when $\tilde{Q}^1+\tilde{Q}^2$ is equal to the zero matrix. Particularly, define functions $\as^1(t)$ and $\as^2(t)$ such that
\begin{align}\label{eq:de}
\as^1(t) = \frac{d\mypi^1(t)}{dt}+ \mypi^1(t)\quad  \mbox{and}\quad \as^2(t) = \frac{d\mypi^2(t)}{dt} + \mypi^2(t),
\end{align}
then we have $\h^1(t) = \as^1(t)^T\tilde{Q}^1\mypi^2(t)$ and $\h^2(t) = \mypi^1(t)^T\tilde{Q}^2\as^2(t)$ since
$$
\as^1(t) \in \argmax_{a\in A^1} \left\{a^T\tilde{Q}^1\mypi^2(t)\right\}\quad\mbox{and}\quad\as^2(t) \in \argmax_{a\in A^2} \left\{\mypi^1(t)^T\tilde{Q}^2a\right\}
$$
by \eqref{eq:brd}. \cite[Section 5]{ref:Harris98} showed that for almost every $t\in[0,\infty)$, $\h^1(t)$ and $\h^2(t)$ are differentiable functions of time and we have\footnote{As pointed out by \cite{ref:Harris98}, the envelope theorem would have led to \eqref{eq:dhdt} if certain smoothness conditions held. Inspired by this intuition, he showed \eqref{eq:dhdt} by using the Taylor series expansion of strategies and by incorporating the closed form solution of the differential equation \eqref{eq:de}.}
\begin{equation}\label{eq:dhdt}
\frac{d\h^1(t)}{dt} = \as^1(t)^T\tilde{Q}^1(\as^2(t)-\mypi^2(t))\quad \mbox{and}\quad \frac{d\h^2(t)}{dt} = (\as^1(t)-\mypi^1(t))^T\tilde{Q}^2\as^2(t).
\end{equation}
Note that \eqref{eq:dhdt} holds irrespective of whether $\tilde{Q}^1+\tilde{Q}^2=O$ or not. Then by definition of $V_H$, as described in \eqref{eq:Harris}, we obtain
\begin{align}
\nonumber\frac{dV_H(\mypi^1(t),\mypi^2(t))}{dt} &= \as^1(t)^T\tilde{Q}^1(\as^2(t)-\mypi^2(t)) + (\as^1(t)-\mypi^1(t))^T\tilde{Q}^2\as^2(t)\\
&= -V_H(\mypi^1(t),\mypi^2(t))+ \as^1(t)^T(\tilde{Q}^1+\tilde{Q}^2)\as^2(t),\label{eq:vhh}
\end{align}
for almost every $t\in[0,\infty)$. Therefore, as can be seen in \eqref{eq:vhh}, the time derivative of $V_H(\mypi^1(t),\mypi^2(t))$ is {\em not} necessarily non-positive for an arbitrary fixed solution $(\mypi^1(t),\mypi^2(t))$. However, when $\tilde{Q}^1+\tilde{Q}^2=O$, the time derivative \eqref{eq:vhh} reduces to
\begin{align}
\frac{dV_H(\mypi^1(t),\mypi^2(t))}{dt} &= - V_H(\mypi^1(t),\mypi^2(t)),\label{eq:dd1}
\end{align}
which is non-positive since $V_H(\mypi^1(t),\mypi^2(t))\geq 0$ when $\tilde{Q}^1+\tilde{Q}^2=O$.\footnote{In particular, $\tilde{Q}^1+\tilde{Q}^2=O$ implies that $\h^2(t) =  \max_{a\in A^2} \left\{\mypi^1(t)^T (-\tilde{Q}^{1}) a\right\}$ and therefore $V_H(\mypi^1(t),\mypi^2(t))$ is bounded from below by
\begin{equation*}
V_H(\mypi^1(t),\mypi^2(t)) \geq \min_{\pi^2\in\Delta( A^2)}\max_{\pi^1\in \Delta(A^1)} \left\{(\pi^1)^T\tilde{Q}^{1} \pi^{2}\right\} - \max_{\pi^1\in\Delta( A^1)}\min_{\pi_2\in \Delta(A^2)} \left\{(\pi^1)^T \tilde{Q}^{1} \pi^{2}\right\}=0,
\end{equation*}
where the equality follows from the Minimax Theorem.} Hence $V_H$ is a Lyapunov function when $\tilde{Q}^1+\tilde{Q}^2=O$.

Our goal here is to modify $V_H(\mypi^1(t),\mypi^2(t))$ to general-sum settings for which the {\em possibility} that
\begin{equation}
\h^1(t) + \h^2(t) < \as^1(t)^T(\tilde{Q}^1+\tilde{Q}^2)\as^2(t)
\end{equation}
poses a challenge as can be seen in \eqref{eq:vhh}. Hence as a candidate Lyapunov function, instead we propose
\begin{align}
V(\mypi^1(t),\mypi^2(t),\Q^1(t),\Q^2(t)) :=\left[\h^1(t) + \h^2(t) - \lambda\|\Q^1(t) + \Q^2(t)\|_{\max}\right]_+,\label{eq:VLyapunov}
\end{align}
where $\lambda>1$ is {\em arbitrary} and is set such that $\gamma\lambda<1$ given the discount factor $\gamma\in(0,1)$. By \eqref{eq:Qdiff}, $V(\cdot)$ reduces to $V_H(\cdot)$ if $\tilde{Q}^1 + \tilde{Q}^2 = O$. When $\|\tilde{Q}^1 + \tilde{Q}^2\|_{\max}>0$, we always have
\begin{equation}
\as^1(t)^T(\tilde{Q}^1+\tilde{Q}^2)\as^2(t) < \lambda\|\tilde{Q}^1 + \tilde{Q}^2\|_{\max}
\end{equation}
since $\lambda >1$. Furthermore, $V(\cdot)$ is a non-negative continuous function by definition irrespective of whether $\tilde{Q}^1 + \tilde{Q}^2 = O$ or not. 

It is also instructive to note that the set $\{x:V(x)=0\}$ is not necessarily the set of equilibria for the continuous-time best response dynamics, \eqref{eq:brd}. However, validity of this candidate function implies that for any solution to the best-response dynamics \eqref{eq:brd}, the sum 
\begin{equation}
\h^1(t)+\h^2(t)= \max_{a\in A^1} \left\{a^T\tilde{Q}^{1} \mypi^{2}(t)\right\}+\max_{a\in A^2}  \left\{\mypi^1(t)^T \tilde{Q}^{2} a\right\}
\end{equation} 
is asymptotically bounded from above by a scaled version of the maximum norm $\|\tilde{Q}^1 + \tilde{Q}^2\|_{\max}$. Later when we focus on discrete-time updates of beliefs on $Q$-functions (at the slow timescale), we will exploit this asymptotic bound to show that the beliefs on $Q$-functions sum to zero asymptotically.

The following lemma shows that $V(\cdot)$ is indeed a Lyapunov function. 

\begin{lemma}\label{lem:Lyapunov}
The candidate function $V(\cdot)$ is a Lyapunov function of the differential inclusion \eqref{eq:brd} for the set $\{x:V(x)=0\}$. In other words, for any trajectory $\x(t)=(\mypi^1(t),\mypi^2(t),\Q^1(t),\Q^2(t))$ of \eqref{eq:brd}, we have
\begin{itemize}
\item $V(\x(t')) < V(\x(t))$ for all $t'>t$ if $V(\x(t))>0$,
\item $V(\x(t')) = 0$ for all $t'>t$ if $V(\x(t))=0$.
\end{itemize} 
\end{lemma}

\begin{proof}
Fix an arbitrary solution to \eqref{eq:brd}, $\x(t)$, and define the function
\begin{equation}
L(t):=\h^1(t)+\h^2(t)-\lambda \|\tilde{Q}^1+\tilde{Q}^2\|_{\max}
\end{equation}
so that $V(\x(t))=[L(t)]_+$. Note that $L(\cdot)$ is absolutely continuous because the solution $(\pi^1(\cdot),\pi^2(\cdot))$ is absolutely continuous, $\max\{\cdot\}$ and addition satisfy Lipschitz condition \cite[Lemma 4.3.2]{ref:Bogachev20book}. Furthermore, the term $\lambda \|\tilde{Q}^1+\tilde{Q}^2\|_{\max}$ does not depend on time. Hence we can compute the time derivative almost everywhere by using \eqref{eq:dhdt}, which leads to 
\begin{align}
\nonumber\frac{dL(t)}{dt} &= - (\h^1(t)+ \h^2(t)) + \as^1(t)^T(\tilde{Q}^1+\tilde{Q}^2)\as^2(t)\\
&< - (\h^1(t)+ \h^2(t)) + \lambda\|\tilde{Q}^1+\tilde{Q}^2\|_{\max}= -L(t),\label{eq:L}
\end{align}
where the strict inequality follows since $\lambda >1$. Therefore, the absolutely continuous $L(t)$ is {\em strictly} decreasing whenever $L(t) \geq 0$. 

On the other hand, by definition, $V(\x(t)) = L(t)$ if $L(t)\geq 0$. Further $V(\x(t))=0$ if $L(t) <0$.  Therefore $V(\x(t))$ is strictly decreasing if, and only if, $V(\x(t))>0$.  Further $V(\x(t))$ remains constant if, and only if, $V(\x(t))=0$. This is irrespective of whether $L(t)<0$ increases or decreases since $L(t)$ is strictly decreasing when $L(t)=0$. More explicitly, the only way $L(t)<0$ becoming $L(t')> 0$ for some $t'>t$ is by crossing zero, that is, if there exists a time $t''\in(t,t')$ such that $L(t'')=0$ since $L(\cdot)$ is a continuous function. However, at that time $t''$, by \eqref{eq:L}  the time derivative is {\em strictly} negative leading it back inside the set $(-\infty,0]$ and preventing an escape to positive values.
\end{proof}

\Cref{lem:Lyapunov} and the stochastic differential inclusion yield that 
\begin{equation}\label{eq:limV}
\lim_{k\rightarrow \infty} V(\hat{\pi}_{k}^1(s),\hat{\pi}_{k}^2(s),\hat{Q}_{k}^1(s),\hat{Q}_{k}^2(s)) = 0,\quad \forall\;s\in S
\end{equation}
and the limit set of the Lyapunov function $\{x:V(x)=0\}$ is given by
$$
\left\{(\pi^1,\pi^2,Q^1,Q^2): \max_{a\in A^1}\{a^TQ^1\pi^2\} + \max_{a\in A^2}\{(\pi^1)^TQ^2a\} - \lambda \|Q^1+Q^2\|_{\max}\leq 0\right\}.
$$

In the following step, we characterize the convergence properties of the beliefs on $Q$-functions by using \eqref{eq:limV}.

\subsection{Step $iii)$ Zooming out to the global dynamics at the slow timescale} \label{sec:conv_q}

We will first show that the sum of the payoff matrices in the auxiliary games, i.e., 
\begin{equation}
\bar{Q}_{k}(s):= \hat{Q}_{k}^1(s)+\hat{Q}_{k}^2(s),
\end{equation}
converges to zero based on the limit set characterization \eqref{eq:limV}, and then we will use this result to show that $\hat{v}_{k}^i(s)$ tracks the saddle point associated with $\hat{Q}_{k}^i(s)$, denoted by 
\begin{align}
&\val^i(\hat{Q}_{k}^i(s)) := \max_{\pi^{i}\in\Delta( A^i)}\min_{\pi^{-i}\in\Delta( A^{-i})}\left\{(\pi^1)^T \hat{Q}_{k}(s)^i\pi^{2}\right\},\quad i=1,2.\label{eq:vsum2}
\end{align}

By definition of the Lyapunov function \eqref{eq:VLyapunov}, we can write \eqref{eq:limV} as
\begin{equation}\label{eq:zeta}
\lim_{k\rightarrow\infty} \left[\bar{v}_{k}(s) - \lambda\|\bar{Q}_{k}(s)\|_{\max}\right]_+ = 0,
\end{equation}
where we define the sum $\bar{v}_k(s):=\hat{v}_{k}^1(s) + \hat{v}_{k}^2(s)$. This implies that the sum $\bar{v}_{k}(s)$ is less than or equal to $\lambda\|\bar{Q}_{k}(s)\|_{\max}$ in the limit. On the other hand, we can bound $\bar{v}_{k}(s)$ from below by $-\lambda\|\bar{Q}_{k}(s)\|_{\max}$ as follows:
\begin{align}
\bar{v}_{k}(s) \geq \hat{\pi}_{k}^1(s)^T\hat{Q}_{k}^1(s)\hat{\pi}_{k}^2(s) + \hat{\pi}_{k}^1(s)^T\hat{Q}_{k}^2(s)\hat{\pi}_{k}^2(s)\geq-\lambda \|\bar{Q}_{k}(s)\|_{\max}\label{eq:lll}
\end{align}
since $\lambda >1$. Combining \eqref{eq:zeta} and \eqref{eq:lll}, we obtain
\begin{equation}\label{eq:lowupp}
-\lambda \|\bar{Q}_{k}(s)\|_{\max} \leq \bar{v}_{k}(s) \leq \lambda\|\bar{Q}_{k}(s)\|_{\max} + \overline{\epsilon}_{k}(s),\;\forall s\in S,\, k\geq 0,
\end{equation}
where $\overline{\epsilon}_{k}(s)$ is an asymptotically negligible error for each $s\in S$. Based on \eqref{eq:lowupp} and \arXiv{\cite[Theorem 5.1]{ref:Sayin20}}{\Cref{lem:aux}}, the following lemma exploits the fact that $r_s^1(a)+r_s^2(a)=0$ for all $(s,a)$, and shows that the auxiliary games become zero-sum in the {\em limit} even though they are not necessarily zero-sum in finite time.

\begin{lemma}\label{lem:zeta}
$\|\bar{Q}_{k}(s)\|_{\max} \rightarrow 0$ as $k\rightarrow\infty$ for each $s$.
\end{lemma}

\begin{proof}
The proof follows from $i)$ writing the sum $\bar{Q}_{k}$ in a recursive form based on the updates of $(\hat{Q}_{k}^1,\hat{Q}_{k}^2)$ and then $ii)$ showing that this recursion satisfies the conditions listed in \arXiv{\cite[Theorem 5.1]{ref:Sayin20}}{\Cref{lem:aux}}. 

{\em Step $a)$} Based on the fact that $r^1(s,a)+r^2(s,a)=0$ for all $(s,a)$, the update of the beliefs on $Q$-functions, as described in \eqref{eq:Qfunc}, yields that 
\begin{align}\label{eq:QQ}
\bar{Q}_{k+1}(s,a) = (1-\bar{\beta}_{k}(s))\bar{Q}_{k}(s,a)+\bar{\beta}_{k}(s)\gamma \sum_{\tilde{s}\in S }p(\tilde{s}|s,a) \bar{v}_{k}(\tilde{s}) ,\quad \forall a.
\end{align}
where $\bar{\beta}_{k}(s) := \indicator{s=s_{k}}\beta_{\#s}\in[0,1]$. Note that without $r^1(s,a)+r^2(s,a)=0$ for all $(s,a)$, there would be an additional constant term in the evolution of the sequence $\{\bar{Q}_{k}(s,a)\}_{k\geq 0}$ preventing its convergence to zero. 

{\em Step $b)$} Based on \eqref{eq:lowupp} and \eqref{eq:QQ}, we can bound $\bar{Q}_{k+1}(s,a)$ as follows
\begin{subequations}\label{eq:Qbarupdate}
\begin{align}
\bar{Q}_{k+1}(s,a) &\leq  (1-\bar{\beta}_{k}(s))\bar{Q}_{k}(s,a)+\bar{\beta}_{k}(s)\left(\bar{\gamma} \max_{s'\in S }\|\bar{Q}_{k}(s')\|_{\max} + \overline{\varepsilon}_{k}\right),\label{eq:mmu}\\
\bar{Q}_{k+1}(s,a) &\geq (1-\bar{\beta}_{k}(s))\bar{Q}_{k}(s,a)-\bar{\beta}_{k}(s)\bar{\gamma} \max_{s'\in S }\|\bar{Q}_{k}(s')\|_{\max},\label{eq:mml}
\end{align}
\end{subequations}
where $\bar{\gamma}:=\gamma\lambda \in (0,1)$ and $\overline{\varepsilon}_{k} = \gamma\max_{s'\in S} \overline{\epsilon}_{s',k}$ is an asymptotically negligible error. 
\arXiv{We emphasize that $\gamma\lambda \in (0,1)$ plays an important role in \cite[Theorem 5.1]{ref:Sayin20}. Based on \Cref{assume:visits,assume:steps}, the step size $\bar{\beta}_{k}(s)\in[0,1]$ satisfies the conditions listed in \cite[Theorem 5.1]{ref:Sayin20}. Furthermore, the iterates are bounded since the stage-payoffs have compact support, $\gamma\in(0,1)$ and the step sizes take values only in $[0,1]$. Therefore, we can invoke \cite[Theorem 5.1]{ref:Sayin20} and complete the proof.}{
We emphasize that $\gamma\lambda \in (0,1)$ plays an important role in resorting to \Cref{lem:aux}. Let $Z:=S\times A$ be the set of all possible state-action pairs. Within the framework of \arXiv{\cite[Theorem 5.1]{ref:Sayin20}}{\Cref{lem:aux}}, we define the vector $y_k \in \mathbb{R}^{|Z|}$ such that its $z$th entry corresponding to state $s$ and the action profile $a$ is defined by $y_{k}[z] := \bar{Q}_{k}(s,a)$, for all $k\geq 0$. This yields that $\|y_k\|_{\infty} = \max_{s\in S }\|\bar{Q}_{k}(s)\|_{\max}$, which is bounded for all $k$ since the stage-payoffs have compact support, $\gamma\in(0,1)$ and the step sizes take values only in $[0,1]$. Based on \Cref{assume:visits,assume:steps}, the step size $\bar{\beta}_{k}(s)\in[0,1]$ satisfies the conditions listed in \Cref{lem:aux}. Therefore, we can invoke \Cref{lem:aux} and obtain $\|y\|_{\infty} \rightarrow 0$ as $k\rightarrow\infty$, and the proof is completed.}
\end{proof}

Since the maximum norm of $\bar{Q}_{k}(s)$ converges to zero by \Cref{lem:zeta}, the upper bound on $\bar{v}_k$, as described in \eqref{eq:lowupp}, implies
\begin{equation}\label{eq:vsum}
\lim_{k\rightarrow\infty} |\bar{v}_{k}(s)|  = 0,\;\forall s\in S. 
\end{equation}
Based on \eqref{eq:vsum} and \Cref{lem:zeta}, the following lemma shows that estimates of continuation payoffs track the minimax values associated with the beliefs on $Q$-functions.

\begin{lemma}\label{lem:track}
$|\hat{v}_{k}^i(s) - \val^i(\hat{Q}_{k}^i(s))| \rightarrow 0$ as $k\rightarrow\infty$ for each $(i,s)$.
\end{lemma}

\begin{proof}
Set $i=1$. Then, the stationary-point inequality says that
\begin{align}
\hat{v}_k^1(s) = \max_{a\in A^1}\{a^T\hat{Q}_k^1(s)\hat{\pi}_k^2(s)\} \geq \val^1(\hat{Q}_k^1(s)) \geq \min_{a\in A^2}\{\hat{\pi}^1(s)^T\hat{Q}_k^1(s)a\}.
\end{align}
Since $\bar{Q}_k(s)=\hat{Q}_k^1(s)+\hat{Q}_k^2(s)$, the right-hand side is bounded from below by
\begin{align}
\min_{a\in A^2}\{\hat{\pi}^1(s)^T\hat{Q}_k^1(s)a\} &\geq \min_{a\in A^2}\{\hat{\pi}^1(s)^T(-\hat{Q}_k^2(s))a\} + \min_{a\in A^2}\{\hat{\pi}^1(s)^T\bar{Q}_k(s)a\}\nonumber\\
&=-\max_{a\in A^2}\{\hat{\pi}^1(s)^T\hat{Q}_k^2(s)a\} + \min_{a\in A^2}\{\hat{\pi}^1(s)^T\bar{Q}_k(s)a\}\nonumber\\
&\geq -\max_{a\in A^2}\{\hat{\pi}^1(s)^T\hat{Q}_k^2(s)a\} - \|\bar{Q}_k(s)\|_{\max}.
\end{align}
Then, the definition of $\hat{v}_k^2(s)$ yields that 
$\hat{v}_k^1(s) \geq \val^1(\hat{Q}_k^1(s)) \geq -\hat{v}_k^2(s) - \|\bar{Q}_k(s)\|_{\max}$.
Since the difference between the first and second terms is bounded from above by the difference between the first and third term, we have
\begin{equation}\label{eq:629}
\hat{v}_k^1(s)+\hat{v}_k^2(s) + \|\bar{Q}_k(s)\|_{\max} \geq \hat{v}_k^1(s)-\val^1(\hat{Q}_k^1(s)) \geq 0.
\end{equation}
Since $\bar{v}_k(s)=\hat{v}_k^1(s)+\hat{v}_k^2(s)$, the left-hand side goes to zero by \Cref{lem:zeta} and \eqref{eq:vsum}. By symmetry, the result can be generalized to $i=2$, which completes the proof.
\end{proof}

Next we introduce the Shapley operator $\mathcal{T}^{i}$ for each $i=1,2$, where
\begin{align}\label{eq:Tcontraction}
(\mathcal{T}^iQ^i)(s,a) := r^i(s,a) + \gamma \sum_{\tilde{s}\in S } p(\tilde{s}|s,a) \val^i(Q^i(\tilde{s})).
\end{align}
\cite{ref:Shapley53} showed that the Shapley operators have contraction property, i.e.,
\begin{equation}\label{eq:Shapley}
\max_{(s,a)\in S\times A } |(\mathcal{T}^{i}Q^i)(s,a)-(\mathcal{T}^i\tilde{Q}^i)(s,a)| \leq \gamma \max_{s\in S }\|Q^i(s)-\tilde{Q}^i(s)\|_{\max}
\end{equation}
because $\gamma\in(0,1)$ and
\begin{equation}\label{eq:valval}
|\val^i(Q^i(s)) - \val^i(\tilde{Q}^i(s))|\leq \|Q^i(s) - \tilde{Q}^i(s)\|_{\max}
\end{equation}
for any $Q^i(s)$ and $\tilde{Q}^i(s)$. Therefore, the Shapley operator $\mathcal{T}^i$ has a \textit{unique} fixed point, denoted by $Q_*^i$, corresponding to the $Q$-functions associated with any stationary equilibrium in the underlying zero-sum stochastic game. Furthermore, $Q_{*}^1(s,a)+Q_{*}^2(s,a)=0$ for all $(s,a)\in S\times A$.

At stage $k$ and state $s\in S $, the update of beliefs on $Q$-functions, e.g., \eqref{eq:Qfunc}, can be written as
\begin{align}\label{eq:hQ}
\hat{Q}_{k+1}^i(s,a) = (1-\bar{\beta}_{k}(s))\hat{Q}_{k}^i(s,a) + \bar{\beta}_{k}(s)\left((\mathcal{T}^i\hat{Q}_{k}^i)(s,a) + \bar{\mathcal{E}}_{k}^i(s,a)\right),
\end{align}
where $\bar{\beta}_{k}(s)=\indicator{s=s_{k}}\beta_{\#s}\in[0,1]$ and $\bar{\mathcal{E}}_{k}^i(s,a):=\gamma\sum_{\tilde{s}\in S}p(\tilde{s}|s,a)[\hat{v}_k^i(\tilde{s})-\val^i(\hat{Q}_{k}^i(\tilde{s}))]$ is an asymptotically negligible error matrix due to the tracking result in \Cref{lem:track}. Based on the contraction property of the Shapley operators, the following lemma characterizes the convergence properties of the iteration \eqref{eq:hQ} by invoking \arXiv{\cite[Theorem 5.1]{ref:Sayin20}}{\Cref{lem:aux}}again.

\begin{lemma}\label{lem:Q}  
$|\hat{Q}_{k}^i(s,a)-Q_{*}^i(s,a)| \rightarrow 0$ as $k\rightarrow\infty$, for each $(i,s,a)$.
\end{lemma}

\begin{proof}
Denote $\widetilde{Q}_{k}^i:= \hat{Q}_{k}^i-Q^i_{*}$. If we subtract the fixed point $Q_{*}^i(s,a)$ from both hand side at \eqref{eq:hQ}, we obtain
\begin{align*}
\widetilde{Q}_{k+1}^i(s,a) = (1-\bar{\beta}_{k}(s))\widetilde{Q}_{k}^i(s,a) + \bar{\beta}_{k}(s)\left((\mathcal{T}^i\hat{Q}_{k}^i)(s,a) - (\mathcal{T}^iQ^i_{*})(s,a) + \bar{\mathcal{E}}_{k}^i(s,a)\right)
\end{align*}
since $\mathcal{T}^iQ_*^i = Q_{*}^i$. Correspondingly, \eqref{eq:Shapley} yields that
\begin{subequations}
\begin{align}
&\widetilde{Q}_{k+1}^i(s,a) \leq (1-\bar{\beta}_{k}(s))\widetilde{Q}_{k}^i(s,a) + \bar{\beta}_{k}(s)\left(\gamma\max_{s'\in S}\|\widetilde{Q}_{k}^i(s')\|_{\max} + \epsilon_{k}^i\right),\\
&\widetilde{Q}_{k+1}^i(s,a) \geq (1-\bar{\beta}_{k}(s))\widetilde{Q}_{k}^i(s,a) + \bar{\beta}_{k}(s)\left(-\gamma\max_{s'\in S}\|\widetilde{Q}_{k}^i(s')\|_{\max} -\epsilon_{k}^i\right),
\end{align}
\end{subequations}
for all $a$, where $\epsilon_{k}:=\max_{(s,a)}|\mathcal{E}_{k}^i(s,a)|$ is an asymptotically negligible error. 
\arXiv{The proof is completed by invoking \cite[Theorem 5.1]{ref:Sayin20}.}
{
Within the framework of \arXiv{\cite[Theorem 5.1]{ref:Sayin20}}{\Cref{lem:aux}}, we define the vector $y_k \in\mathbb{R}^{| Z |}$, where $Z=S\times A$, such that its $z$th entry corresponding to state $(s,a)$ is now defined by $y_{k}[z] := \widetilde{Q}_{k}^i(s,a)$, for all $k$. This yields that $\|y_k\|_{\infty} = \max_{s\in S }\|\widetilde{Q}_{k}^i(s)\|_{\max}$ and $y_k$ is bounded. Based on the contraction property of the Shapley operator, we can invoke \Cref{lem:aux} and obtain $\|y_k\|_{\infty}\rightarrow 0$ as $k\rightarrow\infty$, and the proof is completed.}
\end{proof}

Based on \eqref{eq:valval}, \Cref{lem:track,lem:Q} yield that $|\hat{v}_k^i(s) - \val^i(Q_{*}^i(s))| \rightarrow 0$ as $k\rightarrow\infty$, for all $s\in S$ and $i=1,2$. Therefore the beliefs on the opponent's strategies also converge to an equilibrium of the underlying zero-sum stochastic game. This completes the proof of \Cref{thm:main}. 
\arXiv{The proof of \Cref{cor:dev} is deferred to the extended version \cite{ref:Sayin20}.}{

\subsection{Proof of \Cref{cor:dev}}

The proof follows similar lines with the proof of \Cref{thm:main} except the last step. For example, we now have
\begin{equation}
\limsup_{k\rightarrow\infty} \|\bar{Q}_k(s)\|_{\max} \leq \frac{d}{1-\gamma},\quad \forall s, 
\end{equation}
by \Cref{lem:aux}. Correspondingly, \eqref{eq:lowupp} yields that
\begin{equation}
\limsup_{k\rightarrow\infty} |\bar{v}_k(s)| \leq \frac{\lambda d}{1-\gamma},\quad\forall s. 
\end{equation}
Furthermore, by \eqref{eq:629}, we obtain
\begin{equation}
\limsup_{k\rightarrow\infty}|\hat{v}_{k}^i(s) - \val^i(\hat{Q}_{k}^i(s))| \leq \frac{d(1+\lambda)}{1-\gamma} < \frac{d(1+\gamma)}{\gamma(1-\gamma)},\quad\forall s,
\end{equation}
since $\lambda < 1/\gamma$ by its definition. Invoking \Cref{lem:aux} again completes the proof.
}

\section{Proof of \Cref{cor:main}: Convergence for the Model-free Setting}\label{app:cor:main}

The proof of \Cref{cor:main} follows similar lines with the proof of \Cref{thm:main}. However, we face additional challenges such as the players update the beliefs on the $Q$-functions only for the current state and joint action pair, and they explore by taking any action randomly since they do not know their payoff functions and state transition probabilities. In the following, we focus on how to address these challenges.

{\em Step $i)$ Decoupling the dynamics at the fast timescale.} Similar to \eqref{eq:Qeps}, the update of the belief on $Q$-function at stage $k$ could be written as
\begin{equation}\label{eq:Qeps2}
\hat{Q}_{k+1}^i(s,a) = \hat{Q}_{k}^i(s,a) + \bar{\alpha}_{k}(s)\tilde{\mathcal{E}}_{k}^i(s,a),
\end{equation}
for each $(s,a)$, and the error term $\tilde{\mathcal{E}}_{k}^i(s,a)$ is now defined by
\begin{equation}\label{eq:err2}
\tilde{\mathcal{E}}_{k}^i(s,a) = \indicator{(s,a)=(s_k,a_k)}\frac{\beta_{\#(s,a)}}{\alpha_{\#s}}\left(r_{k}^i + \gamma \hat{v}_{k}^i(s_{k+1}) - \hat{Q}_{k}^i(s,a)\right),
\end{equation}
where $s_k,s_{k+1}$ denote the current and next states, and $a_k$ denotes the current action profile.
The beliefs on the $Q$-functions remain bounded also in the model-free setting. Particularly, for each $s\in S$, we have 
\begin{equation}\label{eq:DD}
\limsup_{k\rightarrow\infty} \|\hat{Q}_k^i(s)\|_{\max} \leq \frac{1}{1-\gamma} \max_{(s',a')\in S\times A} |r^i(s',a')|=:D^i
\end{equation}
independent of the initialization by $\gamma \in (0,1)$ and \Cref{assume:visits,assume:steps}.
 Although we have the ratio $\beta_{\#(s,a)}/\alpha_{\#s}$ in \eqref{eq:err2} instead of $\beta_{\#s}/\alpha_{\#s}$ different from \eqref{eq:err}, the following lemma shows that $\tilde{\mathcal{E}}_{k}^i(s,a)$ is still asymptotically negligible almost surely.

\begin{lemma}\label{lem:error}
Under \Cref{assume:visits,assume:steps,assume:stepFree}, we have $\tilde{\mathcal{E}}_{k}^i(s,a) \rightarrow 0$ almost surely for each $(i,s,a)$ as $k\rightarrow\infty$.
\end{lemma} 

\arXiv{\begin{proof}
The proof is provided in the extended version \cite{ref:Sayin20}.
\end{proof}}
{\begin{proof}
If $|A^1|\times |A^2|=1$, then $\#(s,a)=\#s$ and the result follows from \Cref{assume:steps} since the iterates are bounded. Suppose $|A^1|\times|A^2|>1$. Then, \eqref{eq:DD} yields that 
\begin{equation}
|\tilde{\mathcal{E}}_{k}^i(s,a)| \leq \indicator{(s,a)=(s_k,a_k)}\frac{\beta_{\#(s,a)}}{\alpha_{\#s}} 2D^i,\quad \forall a\in A.
\end{equation}
Therefore, the error term is asymptotically negligible if we have
\begin{equation}\label{eq:infsum}
\sum_{k\geq 0} \prob{\indicator{(s,a)=(s_k,a_k)}\frac{\beta_{\#(s,a)}}{\alpha_{\#s}} > \tau} < \infty,
\end{equation}
for arbitrary $\tau>0$ based on the Borel-Cantelli Lemma.

Note that $\#s$ and $\#(s,a)$ depend on the stage $k$ implicitly and they are random variables. Let us focus on a single term of the sum \eqref{eq:infsum} for fixed number of visits to state $s$, i.e., $\#s = c$:
\begin{align}
\prob{\frac{\beta_{\#(s,a)}}{\alpha_c} > \tau} \leq \prob{\#(s,a)\leq l_c},\label{eq:oneterm} 
\end{align} 
where $l_c := \max\left\{l\in\mathbb{Z}\,|\,l\leq c \mbox{ and }\beta_{l}/\alpha_c>\tau\right\}$, and the inequality follows since $\{\beta_c\}_{c\geq0}$ is monotonically decreasing by \Cref{assume:stepFree}. Without loss of generality, we assume that $\tau>0$ is sufficiently small such that $l_c$ is well-defined.

Next, we will formulate an upper bound on $\prob{\#(s,a)\leq l_c}$. The exploration with probability $\epsilon>0$ ensures that the probability that the joint action $a$ occurs is bounded from below by $\underline{p}:=\epsilon^2/(|A^1||A^2|)>0$ and from above by $\overline{p}:=1-\underline{p}$ since there exists $a'\neq a$ (by $|A^1|\times|A^2|>1$) and the minimum probability that $a'$ occurs is also $\underline{p}$. Therefore, we can bound $\prob{\#(s,a)\leq l_c}$ from above by
\begin{align}
\prob{\#(s,a)\leq l_c} &\leq \sum_{l=0}^{l_c}\binom{c}{l} \overline{p}^l (1-\underline{p})^{c-l}\\
&= (1-\underline{p})^c \sum_{l=0}^{l_c}\binom{c}{l}
\end{align}
since $\overline{p}=1-\underline{p}$. 

\Cref{assume:stepFree} yields that for any given $m\in (0,1)$, there exists $C\in\mathbb{N}$ such that 
$$
\frac{\beta_{\lfloor mc \rfloor}}{\alpha_c} < \tau, \quad\forall c\geq C.
$$
Since $\{\beta_c\}_{c\geq 0}$ is monotonically decreasing, we have 
$$
\frac{\beta_l}{\alpha_c} \leq \frac{\beta_{\lfloor mc \rfloor}}{\alpha_c} < \tau,\quad\forall \ell \geq mc \geq \lfloor mc \rfloor\quad\mbox{and}\quad c\geq C.
$$
This can also be interpreted as for any given $m$, there exists $C$ such that for all $c\geq C$, the ratio $\beta_l/\alpha_c$ can be larger than $\tau$ only if $l < mc$, or equivalently, 
$$
l_c =\max\left\{l\in\mathbb{Z}\,|\,l\leq c \mbox{ and }\beta_{l}/\alpha_c>\tau\right\} \leq mc\quad \forall c>C.
$$

We can pick the arbitrary $m\in(0,1)$ such that $m<1/2$ because if $l_c<c/2$, then $\sum_{l=0}^{l_c}\binom{c}{l}\leq 2^{H(l_c/c)c}$, where $H(p)=-p\log(p) - (1-p)\log(1-p)$ is the entropy function for the Bernoulli distribution $\mathrm{Ber}(p)$, e.g., see \cite[Lemma 16.19]{ref:Flum06book}. Correspondingly, we can choose the arbitrary $m\in (0,1/2)$ such that
\begin{equation}
\xi:=(1-\underline{p})2^{H(m)}<1,
\end{equation} 
because $H(\cdot)$ is a continuous function taking values between zero ($H(0)=H(1)=0$) and one ($H(1/2)=1$), and $(1-\underline{p})\in(0,1)$. Then, we obtain $\prob{\#(s,a)\leq l_c} \leq \xi^c$, for all $c\geq C$, where $\xi\in(0,1)$. Therefore, we have
\begin{equation}
\sum_{c\geq C} \prob{\#(s,a)\leq l_c} \leq \sum_{c\geq C} \xi^{c} <\infty,
\end{equation}
for any $\tau$.
This completes the proof.
\end{proof}}

{\em Step $ii)$ Zooming into the local dynamics at the fast timescale.} \Cref{lem:error} enables us to decouple dynamics across states as in Section \ref{sec:proof}. Given that \player{i} takes $a_k^i$, \eqref{eq:explore} yields that  the update of $\hat{\pi}_k^i$ can be written as
\begin{equation}
\hat{\pi}_{k+1}^i(s) = \hat{\pi}_k^i(s) + \bar{\alpha}_{k}(s)(\mathbb{E}\{a_k^i\} - \hat{\pi}_k^i + \nu_k^i(s))
\end{equation}
where the stochastic approximation error induced by the fixed-probability exploration is given by 
$$
\nu_k^i(s) := a_k^i - \mathbb{E}\{a_k^i\}.
$$
Note that the expectation is taken with respect to the random exploration and $\mathbb{E}\{a_k^i\} = a_*^i (1-\epsilon) + \frac{1}{|A^i|}\mathbf{1}\epsilon$, where $\mathbf{1}$ denotes a vector whose entries are all ones with the associated dimension, since $\mathbb{E}\{u^i\} =  \frac{1}{|A^i|}\mathbf{1}$ if $u^i\sim \mathcal{U}(A^i)$. By its definition, $\{\nu_k^i(s)\}_{k\geq0}$ is a square integrable Martingale difference sequence. Since $\sum_{c=0}^{\infty}\alpha_c^2 < \infty$ by \Cref{assume:stepFree}, the limiting differential inclusion, i.e., the counterpart of \eqref{eq:brd}, is now given by
\begin{align}\label{eq:brdfree}
\frac{d\mypi^i(t)}{dt} = (1-\epsilon)\as^i(t) + \epsilon \bar{u}^i - \mypi^i(t),
\end{align}
due to the exploration and $\bar{u}^i := \frac{1}{|A^i|}\mathbf{1}$, for $i=1,2$. To address this, we modify the Lyapunov function \eqref{eq:VLyapunov} as follows:
\begin{align}
\bar{V}(\mypi^1(t),\mypi^2(t),\tilde{Q}^1,\tilde{Q}^2) :=\left[\h^1(t) + \h^2(t) - \lambda\zeta(\tilde{Q}^1,\tilde{Q}^2)\right]_+,\label{eq:VLyapunovFree}
\end{align}
where $\zeta(\cdot)$ is defined by 
\begin{equation}
\zeta(\tilde{Q}^1,\tilde{Q}^2):= (1-\epsilon)\|\tilde{Q}^1 + \tilde{Q}^2\|_{\max} + \epsilon\|\tilde{Q}^1\|_{\max}+\epsilon\|\tilde{Q}^2\|_{\max},
\end{equation} 
and $\h^i(t)$ is as described in \eqref{eq:hij}. Note that \eqref{eq:VLyapunovFree} reduces to \eqref{eq:VLyapunov} when $\epsilon=0$, i.e., no experimentation. Its validity can be shown as in \Cref{lem:Lyapunov} if we follow the lines in \cite[Section 5]{ref:Harris98} but for the dynamics \eqref{eq:brdfree} instead. For example, we now have
\begin{subequations}
\begin{align}
&\frac{d\h^1(t)}{dt} = \as^1(t)^T\tilde{Q}^1\left((1-\epsilon)\as^2(t) + \epsilon \bar{u}^2 - \mypi^2(t)\right),\\ 
&\frac{d\h^2(t)}{dt} = \left((1-\epsilon)\as^1(t) + \epsilon \bar{u}^1 - \mypi^1(t)\right)^T\tilde{Q}^2\as^2(t),
\end{align}
\end{subequations}
for almost every $t\in[0,\infty)$ and
\begin{align}
\frac{d\h^1}{dt}+\frac{d\h^2}{dt} = &(1-\epsilon)(\as^1)^T(\tilde{Q}^1+\tilde{Q}^2)\as^2 + \epsilon((\as^1)^T\tilde{Q}^1\bar{u}^2 + (\bar{u}^1)^T\tilde{Q}^2\as^2) - (\h^1+\h^2)\nonumber\\
<&\lambda\zeta(\tilde{Q}^1,\tilde{Q}^2) - (\h^1+\h^2)
\end{align}
if $\tilde{Q}^1+\tilde{Q}^2$ is not equal to the zero matrix.
The validity of $\bar{V}(\cdot)$ yields that
\begin{equation}\label{eq:limFree}
\lim_{k\rightarrow \infty} \left[\bar{v}_k(s) - \lambda\left((1-\epsilon)\|\bar{Q}_{k}(s)\|_{\max} + \epsilon \sum_{i=1,2}\|\hat{Q}_k^i(s)\|_{\max}\right)\right]_+ = 0,\quad s\in S.
\end{equation}

{\em Step $iii)$ Zooming out to the global dynamics at the slow timescale.} Let us select the arbitrary $\lambda >1$ such that $\lambda(1-\epsilon)<1$ in addition that $\lambda\gamma<1$. By \eqref{eq:limFree}, as a counterpart of \eqref{eq:lowupp}, we have
\begin{equation}\label{eq:lowupp2}
-\|\bar{Q}_{k}(s)\|_{\max} \leq \bar{v}_k(s)\leq \|\bar{Q}_{k}(s)\|_{\max} + \overline{\epsilon}_{k}(s), \quad \forall s\in S, k\geq 0,
\end{equation}
but with an error term satisfying $\limsup_{k\rightarrow\infty} |\overline{\epsilon}_{k}(s)| \leq \lambda \epsilon (D^1+D^2)$ (and $D^i$ is as described in \eqref{eq:DD}) for each $s\in S$. Furthermore, we need to consider stochastic approximation error terms induced by sampling the underlying state transition probabilities, which is given by
\begin{equation}\label{eq:omega}
\omega_{k}^i(s,a) := \gamma\hat{v}_{k}^i(s_{k+1}) - \gamma\sum_{\tilde{s}\in S }\hat{v}_{k}^i(\tilde{s}) p(\tilde{s}|s,a),
\end{equation}
for each $i=1,2$, and it is a square integrable Martingale difference sequence by its definition. Note that the sum of approximation errors $\bar{\omega}_{k} := \omega_{k}^1 +\omega_{k}^2$ is also a square integrable Martingale difference sequence. Then, the proof follows after some algebra similar to the ones in Step $iii)$ in the proof of \Cref{thm:main}. 

\arXiv{The relevant technical details are deferred to the extended version \cite{ref:Sayin20}. This completes the proof of \Cref{cor:main}.}{Based on \eqref{eq:lowupp2} and \eqref{eq:omega}, the evolution of $\bar{Q}_{s,k}$ satisfies
\begin{align*}
&\bar{Q}_{k+1}(s,a)\leq (1-\bar{\beta}_{(s,a),k}) \bar{Q}_{k}(s,a) + \bar{\beta}_{(s,a),k}\left(\gamma\max_{s'\in S}\|\bar{Q}_{k}(s')\|_{\max}+ \overline{\varepsilon}_k +\bar{\omega}_{k}(s,a)\right),\\
&\bar{Q}_{k+1}(s,a)\geq (1-\bar{\beta}_{(s,a),k}) \bar{Q}_{k}(s,a) + \bar{\beta}_{(s,a),k}\left(-\gamma\max_{s'\in S}\|\bar{Q}_{k}(s')\|_{\max}+\bar{\omega}_{k}(s,a)\right),
\end{align*}
where $\bar{\beta}_{(s,a),k}:=\indicator{(s,a)=(s_k,a_k)}\beta_{\#(s,a)}$ and $\overline{\varepsilon}_k:=\gamma\max_{s'\in S}\overline{\epsilon}_{k}(s')$, as a counterpart of \eqref{eq:Qbarupdate}.\footnote{Note that independent zero-mean compact support random perturbations on the stage-payoffs will not constitute a technical challenge here because we can incorporate the perturbation to the stochastic approximation error.} Since $\sum_{c=0}^{\infty}\beta_c^2<\infty$ by \Cref{assume:stepFree} and $\limsup_{k\rightarrow\infty}|\overline{\varepsilon}_k|\leq \gamma \lambda \epsilon D$, where $D=D^1+D^2$, we can invoke \arXiv{\cite[Theorem 5.1]{ref:Sayin20}}{\Cref{lem:aux}} and obtain
\begin{equation}\label{eq:barQfree}
\limsup_{k\rightarrow\infty} \|\bar{Q}_{k}(s)\|_{\max} \leq \frac{\gamma\lambda\epsilon D}{1-\gamma} < \frac{\epsilon D}{1-\gamma},\quad \forall s\in S
\end{equation}
since $\lambda \gamma <1$.

On the other hand, the inequalities \eqref{eq:lowupp2} \eqref{eq:barQfree} yield that
\begin{align}
\limsup_{k\rightarrow\infty} |\bar{v}_k(s)| &\leq \limsup_{k\rightarrow\infty} \|\bar{Q}_k(s)\|_{\max} + \lambda\epsilon D\nonumber\\
&\leq \epsilon D \left(\frac{1}{1-\gamma} + \frac{1}{\gamma}\right)\label{eq:barvfree}
\end{align}
since $\lambda<1/\gamma$ by its definition.
Therefore, integrating \eqref{eq:barQfree} and \eqref{eq:barvfree} into \eqref{eq:629}, we obtain 
\begin{equation}\label{eq:vtrack}
\limsup_{k\rightarrow\infty} |\hat{v}_k^i(s) - \val^i(\hat{Q}_{k}^i(s))| < \epsilon D\,\left(\frac{2}{1-\gamma}+\frac{1}{\gamma}\right). 
\end{equation}

Based on the definition of the Shapley operator \eqref{eq:Tcontraction}, the evolution of $\widetilde{Q}_{k}^i(s,a) = \hat{Q}_{k}^i(s,a)-Q_{*}^i(s,a)$ can be written as
\begin{align}
\widetilde{Q}_{k+1}^i(s,a) =& (1-\bar{\beta}_{(s,a),k})\widetilde{Q}_{k}^i(s,a)\nonumber  \\
&+ \bar{\beta}_{(s,a),k}\left[(\mathcal{T}^i\hat{Q}_{k}^i)(s,a)- (\mathcal{T}^iQ_{*}^i)(s,a) + \hat{\mathcal{E}}_{k}^i(s,a) + \omega_{k}^i(s,a)\right],
\end{align}
where $\hat{\mathcal{E}}_{k}^i(s,a):=\gamma\sum_{\tilde{s}\in S}p(\tilde{s}|s,a)[\hat{v}_k^i(\tilde{s}) - \val^i(\hat{Q}_{k}^i(\tilde{s}))]$ is asymptotically bounded, i.e., 
\begin{equation}
\limsup_{k\rightarrow\infty}\|\hat{\mathcal{E}}_{s,k}^i[a]\|_{\max} \leq \gamma\epsilon D\,\left(\frac{2}{1-\gamma}+\frac{1}{\gamma}\right)
\end{equation} 
due to the tracking result \eqref{eq:vtrack}.\footnote{Independent zero-mean compact support random perturbations on the stage-payoffs will again not constitute a technical challenge here because we can incorporate them to the stochastic approximation error.}
Following the lines in \Cref{lem:Q}, we can again resort to \arXiv{\cite[Theorem 5.1]{ref:Sayin20}}{\Cref{lem:aux}} based on the contraction property of the Shapley operator and obtain \eqref{eq:Qresult} because
\begin{equation}\label{eq:Qtrack}
\limsup_{k\rightarrow\infty} \|\hat{Q}_{k}^i(s)- Q_{*}^i(s)\|_{\max} \leq \frac{\gamma\epsilon D}{1-\gamma}\,\left(\frac{2}{1-\gamma}+\frac{1}{\gamma}\right)= \epsilon D \, \frac{1+\gamma}{(1-\gamma)^2},\quad\forall s.
\end{equation}
Furthermore, based on \eqref{eq:valval} and \eqref{eq:vtrack}, \eqref{eq:Qtrack} yields that
\begin{align}
\limsup_{k\rightarrow\infty} &|\hat{v}_k^i(s) - \val^i(Q_{*}^i(s))| \nonumber\\
&\leq \limsup_{k\rightarrow\infty} |\hat{v}_k^i(s) - \val^i(\hat{Q}_{k}^i(s))|  + \limsup_{k\rightarrow\infty} |\val^i(\hat{Q}_{k}^i(s))-\val^i(Q_{*}^i(s))|\nonumber\\
&\leq \limsup_{k\rightarrow\infty} |\hat{v}_k^i(s) - \val^i(\hat{Q}_{k}^i(s))|  + \limsup_{k\rightarrow\infty} \|\hat{Q}_{k}^i(s)- Q_{*}^i(s)\|_{\max} \nonumber\\
&\leq \epsilon D\,\left(\frac{2}{1-\gamma}+\frac{1}{\gamma}\right) + \epsilon D \, \frac{1+\gamma}{(1-\gamma)^2} = \epsilon D \, \frac{1+\gamma}{\gamma(1-\gamma)^2},\quad\forall s.\label{eq:vfinal}
\end{align}

Finally, 
to characterize the convergence properties of the beliefs on the opponent strategies, we will first make explicit the dependence of $Q$-function and value function, respectively, as described in \eqref{eq:q} and \eqref{eq:vis}, on the opponent strategy. For example, they are now denoted by $Q^i(s,a\,|\,\pi^{-i})$ and $v^i(s\,|\,\pi^{-i})$ with slight abuse of notation given that the opponent plays according to $\pi^{-i}$. Furthermore, we pick \player{1} as the typical player. For all $s$ and $k$, we have
\begin{align}
v^1(s\,|\,\hat{\pi}_k^2) - \hat{v}_k^1(s) &= \max_{\mu^1} (\mu^1)^TQ^1(s,\cdot\,|\,\hat{\pi}_k^2)\hat{\pi}_k^2(s) - \max_{\mu^1} (\mu^1)^T\hat{Q}^1_k(s)\hat{\pi}_k^2(s)\nonumber\\
&\leq \|Q^1(s,\cdot\,|\,\hat{\pi}_k^2) -\hat{Q}^1_k(s) \|_{\max}\nonumber\\
&\leq \|Q^1(s,\cdot\,|\,\hat{\pi}_k^2) -Q_*^1(s) \|_{\max} + \|\hat{Q}^1_k(s) -Q_*^1(s)\|_{\max}\label{eq:vv1}
\end{align} 
where the last line follows from the triangle inequality after we add and subtract $Q_*^1(s)$. We have already characterized the convergence properties of the second term on the right-hand side in \eqref{eq:Qtrack}. On the other hand, the definitions of $Q^1(s,a\,|\,\hat{\pi}_k^2)$ and $Q_*^1(s)$ yield that
\begin{equation}
\|Q^1(s,\cdot\,|\,\hat{\pi}_k^2) -Q_*^1(s) \|_{\max}\leq \gamma \max_{\tilde{s}}|v^1(\tilde{s}\,|\, \hat{\pi}_k^2) - v_*^1(\tilde{s})|,\quad \forall s.\label{eq:vv2}
\end{equation}
On the other hand, by the definition of the equilibrium value function, we have
\begin{align}\label{eq:vv3}
0\leq v^1(s\,|\, \hat{\pi}_k^2) - v_*^1(s) = v^1(s\,|\,\hat{\pi}_k^2) - \hat{v}_k^1(s) - (v_*^1(s) - \hat{v}_k^1(s)),
\end{align}
where we have already characterized the convergence properties of the second term on the right-hand side in \eqref{eq:vfinal} since $v_*^i(s) = \val^i(Q_*^i(s))$ for all $s$ while the first term is bounded from above as described in \eqref{eq:vv1} and \eqref{eq:vv2}. Therefore, \eqref{eq:vv1}, \eqref{eq:vv2}, and \eqref{eq:vv3} lead to
\begin{align}
\max_{\tilde{s}}|v^1(\tilde{s}\,|\, \hat{\pi}_k^2) - v_*^1(\tilde{s})| \leq &\gamma \max_{\tilde{s}}|v^1(\tilde{s}\,|\, \hat{\pi}_k^2) - v_*^1(\tilde{s})| + \max_{\tilde{s}}\|\hat{Q}^1_k(\tilde{s}) -Q_*^1(\tilde{s})\|_{\max}\nonumber\\
&+\max_{\tilde{s}}|\hat{v}_k^1(\tilde{s})-v_*^1(\tilde{s})|.
\end{align}
By the limit characterizations \eqref{eq:Qtrack} and \eqref{eq:vfinal}, and the symmetry across players, we obtain
\begin{align}
\limsup_{k\rightarrow\infty} \max_s |v^i(s\,|\, \hat{\pi}_k^{-i}) - v_*^i(s)| \leq \epsilon D \, \frac{1+\gamma}{(1-\gamma)^3}\left(1+\frac{1}{\gamma}\right)=\epsilon D \, \frac{(1+\gamma)^2}{\gamma(1-\gamma)^3}.\label{eq:vVv}
\end{align} 
This is important because for any $\pi^1$, we have
\begin{align}
U^1(\pi^1,\hat{\pi}^2_k) - U^1(\hat{\pi}_k^1,\hat{\pi}_k^2) &= U^1(\pi^1,\hat{\pi}^2_k) + U^2(\hat{\pi}_k^1,\hat{\pi}_k^2)\nonumber\\
&\leq \mathbb{E}_{s\sim p_o}\{v^1(s\,|\,\hat{\pi}_k^2)\} + \mathbb{E}_{s\sim p_o}\{v^2(s\,|\,\hat{\pi}_k^1)\}\nonumber\\
&= \mathbb{E}_{s\sim p_o}\{v^1(s\,|\,\hat{\pi}_k^2)-v_*^1(s)\} + \mathbb{E}_{s\sim p_o}\{v^2(s\,|\,\hat{\pi}_k^1)-v_*^2(s)\}\nonumber\\
&\leq \max_{s}|v^1(s\,|\,\hat{\pi}_k^2)-v_*^1(s)| + \max_{s}|v^2(s\,|\,\hat{\pi}_k^1)-v_*^2(s)|
\end{align}
since $v_*^1(s)+v_*^2(s) = 0$ for all $s$. Combined with \eqref{eq:vVv}, this completes the proof of \Cref{cor:main}.
}

\section{An Illustrative Example}\label{sec:example}

In this section, we examine our fictitious play dynamics numerically in a zero-sum stochastic game whose configuration is selected arbitrarily. For example, there are $3$ states, players have $4$ actions per state, and the discount factor $\gamma=0.8$.  State transition probabilities and stage payoffs are chosen randomly in a way that players can have preferences over the states so that they would face the trade-off between current stage payoff and the continuation payoffs. We set the step sizes as $\alpha_c = 1/(1+c)^{0.51}$ and $\beta_c = 1/(1+c)$ such that they would satisfy both \Cref{assume:steps,assume:stepFree}. Furthermore in the model-free setting, players take a random action with probability $0.02$ in order to learn the unknown state transition probabilities associated with each action. 

\begin{figure}[t!]
\centering
\begin{subfigure}[b]{.495\textwidth}
\centering
\includegraphics[width=\textwidth]{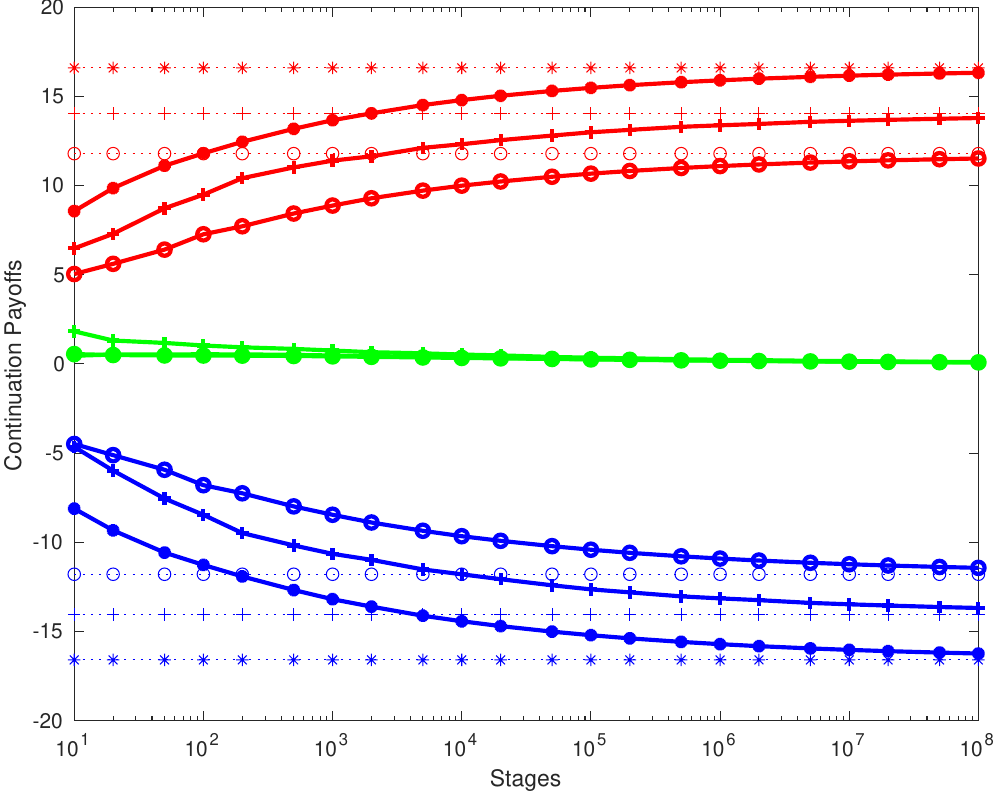}
\caption{Model-based Case}\label{fig:plot_based}
\end{subfigure}
\hfill
\begin{subfigure}[b]{.495\textwidth}
\centering
\includegraphics[width=\textwidth]{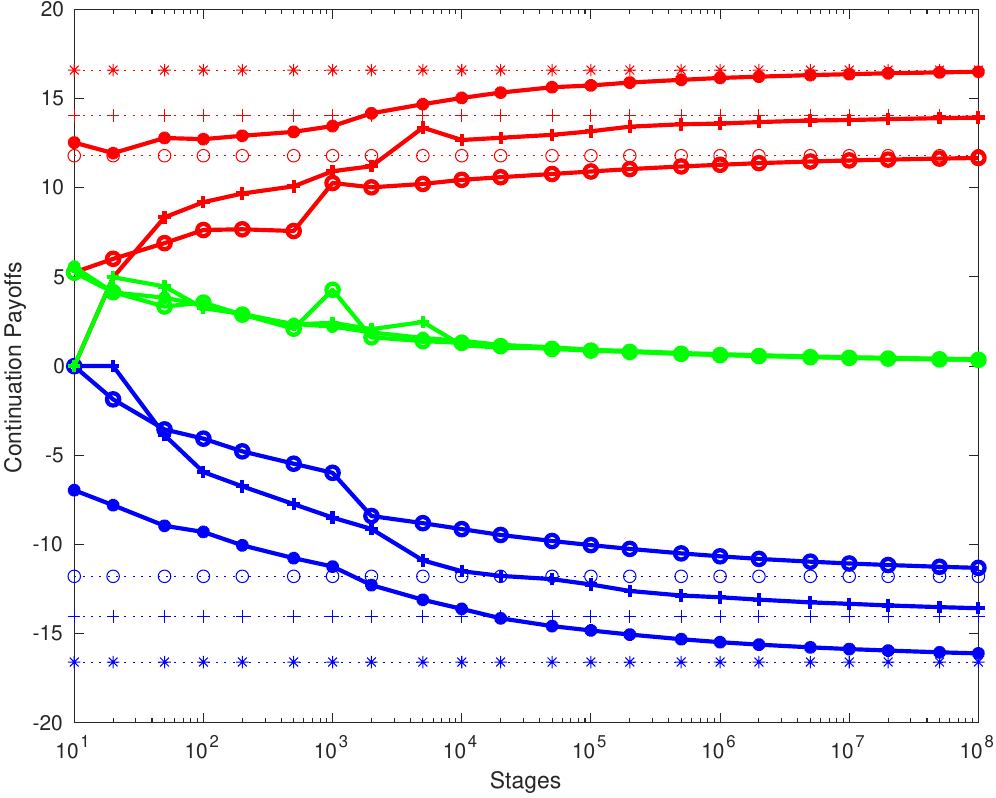}
\caption{Model-free Case}\label{fig:plot_free}
\end{subfigure}
\caption{Evolution of continuation payoff (or value function) estimates $\{\hat{v}_{s,k}^1,\hat{v}_{s,k}^2\}_{s\in S}$ and $\bar{v}_{s,k}=\hat{v}_{s,k}^1+\hat{v}_{s,k}^2$, respectively, converging to positive values, negative values, and zero in \textbf{\textcolor{red}{red}}, \textbf{\textcolor{blue}{blue}}, and \textbf{\textcolor{green}{green}}. Color available in the online version. The dotted lines  {\color{red}\dotted}/{\color{blue}\dotted} denote the actual Nash equilibrium values at each state. The horizontal axis is logarithmic with markers at instances $10,20,50,100,200,\ldots$, in this order.}
\end{figure}

In \Cref{fig:plot_based,fig:plot_free}, we plot the evolution of the continuation payoff estimates of both player for each state in comparison to the equilibrium values, respectively, in model-based and model-free settings. We also plot the sum of continuation payoffs to observe its expected convergence to zero. As expected, we have observed that the estimates of the continuation payoffs converge to the minimax values of each state in the stochastic game while the convergence is relatively slower and more noisy in the model-free setting.

\section{Concluding Remarks}\label{sec:conclusion}

We presented fictitious play dynamics for stochastic games and analyzed its convergence properties in zero-sum games. In the dynamics presented, players form a belief not only on opponent (stationary) strategy but also on the associated $Q$-functions and update them based on the actions taken by the opponent. The update of beliefs on $Q$-functions evolves at a slower timescale compared to the evolution of beliefs on strategies. 

In order to show the convergence of the dynamics, we first approximated the dynamics via a certain differential inclusion at the timescale of the fast update and formulated a novel Lyapunov function for it in order to characterize the limiting behavior of the fast update. Then we used this characterization accompanied with certain contraction arguments at the timescale of the slow update in order to show the almost sure convergence of the dynamics. In particular, we showed that beliefs on strategies and $Q$-functions, respectively, converge to a stationary equilibrium and the corresponding $Q$-functions in the model-based and model-free settings provided that each state is visited infinitely often.

Some of the future research directions include the analyses of this framework $(i)$ in other classes of games, e.g., identical-interest games or zero-sum games with more than two players; $(ii)$ without the conditions on visiting each state infinitely-often, e.g., in terms of self-confirming equilibrium, as studied in \cite{ref:Fudenberg95} for learning in extensive-form game; $(iii)$ with function approximation to address computational challenge due to large state and action spaces; and $(iv)$ with non-asymptotic convergence guarantees.   

\bibliographystyle{siamplain}
\bibliography{mybibfile}
\end{document}